\documentclass[fleqn]{2020SCGE}
\usepackage{graphicx}
\graphicspath{{Figures/}}
\usepackage{comment}
\usepackage{footnote} 
\usepackage{float} 
\begin{document}

\ensubject{Theoretical Physics}

\ArticleType{Article} 
\Year{2023}
\Month{September}
\Vol{66}
\No{9}
\DOI{10.1007/s11433-022-2118-5}
\ArtNo{290411}
\ReceiveDate{December 13, 2022}
\AcceptDate{April 17, 2023}
\OnlineDate{August 7, 2023} 

\title{On the gauge dependence of scalar induced secondary gravitational waves during radiation and matter domination eras}{On the gauge dependence of scalar induced secondary gravitational waves during radiation and matter domination eras}
 
\author[1]{Arshad Ali}{}
\author[1,2,3]{Ya-Peng Hu}{}
\author[4]{Mudassar Sabir}{}
\author[1]{Taotao Sui}{}

\AuthorMark{A. Ali} 
\AuthorCitation{A. Ali, Y.-P. Hu, M. Sabir, and T. Sui}

\address[1]{College of Physics, Nanjing University of Aeronautics and Astronautics, Nanjing 211106, China}
\address[2]{Key Laboratory of Aerospace Information Materials and Physics (NUAA), MIIT, Nanjing 211106, China}
\address[3]{Center for Gravitation and Cosmology, College of Physical Science and Technology, Yangzhou University, Yangzhou 225009, China}
\address[4]{School of Geophysics and Geomatics, China University of Geosciences, Wuhan 430074, China}
 
\abstract{We revisit the vital issue of gauge dependence in the scalar-induced secondary gravitational waves (SIGWs), focusing on the radiation domination (RD) and matter domination (MD) eras.  The energy density spectrum is the main physical observable in such induced gravitational waves. For various gauge choices, there has been a divergence in the energy density, $\Omega_\text{GW}$, of SIGWs. We calculate SIGWs in different gauges to quantify this divergence to address the gauge-dependent problem. In our previous studies, we had found that the energy density  diverges in the polynomial power of conformal time (e.g., $\eta^6$ in uniform density gauge).
 We try to fix this discrepancy by adding
 a counter-term that removes the fictitious terms in secondary tensor perturbations.
We graphically compare the calculations in various gauges  and also comment on the physical origin of the observed gauge dependence.}

\keywords{scalar induced gravitational waves, gauge transformation, cosmology}

\PACS{04.30.-w, 98.80.Cq, 98.80.-k}

\maketitle

\vspace*{1mm}
\begin{multicols}{2}
\section{Introduction}\vspace*{-1mm}
In 2015, Advanced LIGO improved the first network of advanced detectors to be significantly more sensitive to measuring GWs \cite{TheLIGOScientific:2014jea, TheVirgo:2014hva, Affeldt:2014rza, Aso:2013eba,Lin:2018azu,Ghayour:2016fqm,Wang:2021iwp,Pandey:2020gjy,Ghayour:2018tfz,Gwak:2017zwm}. The typical sources for studying the very early universe by cosmological GWs include phase transitions, which lead to the collisions of bubbles or the formation of cosmic strings, resonances during reheating, so-called primordial GWs (quantum fluctuations during inflation), and GWs induced by large primordial fluctuations \cite{Caprini:2018mtu}.  The production of scalar induced gravitational waves (SIGWs) occurs when the large primordial fluctuations re-enter the
\Authorfootnote

\noindent horizon sometime between inflation and the Big Bang  Nucleosynthesis, which is a promising prospect. Because we do not have any evidence of the universe's content or the expansion history at that time, SIGWs allow access to the latter stages of inflation and carry information on the primordial universe's content. Future data on the primordial power spectrum acquired by SIGWs will complement those from other probes, such as spectral distortions \cite{Chluba:2019kpb, Kite:2020uix} in the multimessenger cosmology epoch \cite{Unal:2020mts}.

In addition to SIGWs, there are also GWs with cosmological origins, including primordial GWs derived from inflation and GWs, generated from a cosmic phase transition \cite{Ananda:2006af, Baumann:2007zm, Saito:2008jc, Sabir:2019eut, Sabir:2019xwk, Malik:2023zlb,Sabir:2019wel, Zhang:2020ptw,
Yi:2018gse,
Sathyaprakash:2009xs, Saito:2009jt, Bugaev:2009zh,
	Bugaev:2010bb, Alabidi:2012ex,
	Inomata:2016rbd, Orlofsky:2016vbd,  Garcia-Bellido:2017aan,
	Cheng:2018yyr, Cai:2018dig,  Kohri:2018awv,  Espinosa:2018eve, Lu:2019sti,Cai:2019cdl}.
Even though the primordial GWs are too small to be observed by third-generation ground-based and space-based GW observatories, SIGWs can have peak frequencies as low as nanohertz or millihertz, making them detectable by future space-based GW observatories,  including LISA \cite{Danzmann:1997hm,Audley:2017drz}, TianQin  \cite{Luo:2015ght} and Taiji \cite{Hu:2017mde}, and PTA observations like SKA \cite{Moore:2014lga,Janssen:2014dka}.
Tomita \cite{10.1143/PTP.37.831}  was  the first to note that density fluctuations can induce GWs. Later, in a dust-dominated universe, they were rediscovered in refs. \cite{Matarrese:1992rp,Matarrese:1993zf} when studying second-order cosmological perturbations. 

In contrast to the first-order perturbations, the second-order tensor perturbations induced by scalar perturbations are usually considered to have gauge dependence.
Therefore, the secondary SIGWs may differ depending on gauge choice  \cite{Arroja:2009sh,Hwang:2017oxa,Tomikawa:2019tvi,Inomata:2019yww,Yuan:2019fwv,DeLuca:2019ufz,Lu:2020diy,Chang:2020iji, Nakamura:2011xy,Chang:2020mky},
 despite having many gauge-invariant tensor perturbations at second order  \cite{Bruni:1996im,Mollerach:2003nq,Bartolo:2004ty,Nakamura:2004rm,Nakamura:2006rk,Malik:2008im,Arroja:2009sh,Domenech:2017ems,Blaut:2019fxb,DeLuca:2019ufz,Chang:2020tji,Domenech:2020xin}.
We must determine the secondary tensor perturbations in different gauges for these reasons.  However, SIGW production was typically discussed using the Poisson gauge  \cite{Kohri:2018awv,Ananda:2006af,Bugaev:2009zh,Baumann:2007zm}. Thus, it is vital to examine SIGWs in other gauges. In RD, the energy densities of SIGWs in the Poisson, the TT, and the uniform curvature gauge were identical \cite{Yuan:2019fwv,DeLuca:2019ufz,Inomata:2019yww}.
The energy density of SIGWs in  the TT gauge during RD and MD was examined in \cite{Inomata:2019yww}.
As a general background, the SIGWs have been computed in the Poisson, comoving, and uniform curvature gauges for RD with $w=1/3$ and MD with $w=0$  \cite{Tomikawa:2019tvi}.

Moreover, SIGWs can be measured by analyzing the energy density spectrum \cite{Allen:1997ad}.
The gauge-dependence of SIGWs spectrum has been investigated in refs. \cite{Inomata:2019yww,Yuan:2019fwv,DeLuca:2019ufz}. The calculations in various gauges do not coincide and have risen to confusing statements in the literature \cite{Inomata:2019yww,DeLuca:2019ufz,Tomikawa:2019tvi,Hwang:2017oxa,Giovannini:2020qta}. As pointed out in ref. \cite{Giovannini:2020qta}, the gauge-dependence problem may arise due to the fictitious tensor perturbations upon gauge-fixing. There might be another subtle cause of this discrepancy due to the definition of the physical observable of gravitational waves in the synchronous frame \cite{DeLuca:2019ufz}.
Recently, to study the relation of  SIGWs in different gauges, a framework has been presented \cite{Lu:2020diy,Ali:2020sfw}. The physical behavior of the kernel functions to determine the energy density of SIGWs has not been discussed in depth in the previous studies  \cite{DeLuca:2019ufz, Domenech:2017ems,Gong:2019mui,Bruni:1996im,Matarrese:1997ay,Malik:2008im,Nakamura:2006rk,Chang:2020tji, Lu:2020diy}. The proposed study intends to investigate the gauge dependence of SIGWs in the Poisson gauge, the TT gauge, the comoving orthogonal gauge, the uniform curvature gauge, the total matter gauge, the uniform density gauge, and the uniform expansion gauge in both RD and MD.

 However,  a significant change in the energy density power spectrum of the second-order tensor perturbations induced in an MD era was discussed in ref. \cite{Hwang:2017oxa}. More characteristics of the gauge dependence of the SIGWs have been examined in refs. \cite{Matarrese:1997ay,Arroja:2009sh,Gong:2019mui,Tomikawa:2019tvi}.  The tensor perturbations can be divided into two parts in the context of scalar-induced secondary tensor perturbations. On the one side, these perturbations freely propagate tensor perturbations by following the equation of motion without any source. These kinds of tensor perturbations are widely considered gravitational waves, and the time dependence of these gravitational waves can be read as $h_{ij} \propto \sin(k\eta)\, \text{or}\, \cos(k\eta)$. Despite coupling with scalar perturbations at the production time, they eventually decouple and propagate freely. They no longer depend on the gauge because they are independent of the scalar perturbation once they have decoupled.

 On the other side, the secondary tensor perturbation couples with the scalar perturbations, where the freely propagating secondary tensor perturbations are solely contained until they decouple from the scalar perturbations. Since the scalar perturbations control these kinds of tensor perturbations, the time dependence of these tensor perturbations inherits those of the scalar perturbations. It is to be noted that the gauge dependence appears only in these sorts of tensor perturbations. In literature, in many references, these kinds of tensor perturbations are also usually called gravitational waves. In contrast, to distinguish these two kinds of tensor perturbations in this paper, we call exclusively the freely propagating tensor perturbations or free gravitational waves.

In ref. \cite{DeLuca:2019ufz}, it was claimed that the power spectrum of the energy density of the freely propagating secondary tensor perturbations investigated in the TT gauge is reduced compared with that investigated in the Poisson gauge. Nevertheless, as mentioned above,  one does not envision that the induced secondary GWs depend on the gauge choices.
The gauge independence of the induced secondary GWs can also be anticipated from the coincidence of the
GWs calculated in the Poisson gauge and the flat
gauge, given in ref. \cite{Tomikawa:2019tvi} (see the case of $w>0$ there).

In this study, we  reconsider the subhorizon SIGWs in different popular gauges. In particular, we emphasize on a detailed study of the second-order tensor perturbations generated by linear scalar perturbations in an expanding spacetime containing either RD or MD. More specifically, the paper aims to address the important issue
of the gauge dependence of such tensor perturbations. This problem is highly relevant,
as it is still actively discussed in the literature and has not been properly solved yet, so far as we are aware.
Recent studies such as refs. \cite{Hwang:2017oxa, DeLuca:2019ufz, Lu:2020diy, Tomikawa:2019tvi, Ali:2020sfw} address this problem. By these studies, one  could not obtain a physically meaningful discussion on whether the secondary tensor perturbations induced by (the quadratic of) the first-order scalar perturbations are gauge-invariant.
Besides, there have been discrepancies in the previous studies of SIGWs  \cite{Hwang:2017oxa,Tomikawa:2019tvi,DeLuca:2019ufz, Domenech:2017ems,Gong:2019mui,Bruni:1996im,Matarrese:1997ay,Malik:2008im,Nakamura:2006rk,Chang:2020tji, Lu:2020diy,Ali:2020sfw}. At a late time, the energy density increased as $\eta^n$ in different gauges. Here, $\eta$ represents the conformal time of the universe, and $n$ is an integer. When $\eta\rightarrow\infty$, this strange result in various gauges might indicate a breakdown of the perturbation theory.

However, one interesting finding of the previous investigations is a substantial enhancement of the energy density spectrum during the MD era. Nevertheless, the relevant modes do not redshift as expected for RD. Also, a strong gauge dependence of these secondary induced tensor perturbations was demonstrated in ref. \cite{Hwang:2017oxa}. In addition, the authors (one of us, AA)  in refs. \cite{Lu:2020diy,Ali:2020sfw},  presented a relationship of SIGWs in various gauges. However, the different kernel functions were evaluated without common terms in various gauges during RD. The authors claimed that the energy density is increasing as $\eta^2$ and $\eta^6$ in different gauges, which is actually divergent \cite{Lu:2020diy}.
While, during MD, from the kernels, it seems that the energy density is also in increasing or decreasing modes \cite{Ali:2020sfw}. Furthermore, it extracted some oscillating terms sin$x$ and cos$x$, which are uncommon in all gauges, and presented physically meaningful contributions to SIGWs. For instance, the case of MD lacks convincing from a physical point of view.
Indeed, in previous literature, it is assumed that freely propagating tensor perturbations may contribute to SIGWs during MD. Since these studies mentioned above are flawed during RD and MD.
However, it is crucial to eliminate the terms that cause the discrepancy in different popular gauges.

In our proposal,  we try to fix the discrepancies that occur in the previous studies. However, in the presented gauge independence framework, real and effective sources are called fictitious terms by following the refs. \cite{Nakamura:2020pre,Bruni:1996im,Malik:2008im,Sonego:1997np,Matarrese:1997ay,Chang:2020tji},  we introduce a counter term that removes the fictitious one instead of directly removing it by hand in secondary tensor perturbations. We are now using only the gauge invariant variables, following the refs. \cite{Nakamura:2020pre,Bruni:1996im,Malik:2008im,Sonego:1997np,Matarrese:1997ay,Chang:2020tji} thereby getting rid of the gauge-dependent fictitious terms. We carefully revisit the gauge dependence problem in SIGWs by explicitly calculating  $\Omega_{\text{GW}}$ in seven different gauges.
Thus in the subhorizon modes, the observable energy density spectrum of the SIGWs is the same in seven gauges, in contrast to refs. \cite{Hwang:2017oxa, DeLuca:2019ufz, Lu:2020diy, Tomikawa:2019tvi, Ali:2020sfw}.
 In addition, we make a clear
distinction between scalar-induced secondary tensor perturbations and SIGWs because of
the mixing and coupling of tensor and scalar perturbations. Therefore, further, we recognize the oscillations $\sin{x}$ and $\cos{x}$ in scalar-induced secondary tensor perturbations as
SIGWs during RD and MD.
Basically, in the derivation of scalar-induced secondary tensor
perturbations in different gauges, the physical interpretation is used to identify SIGWs.

 In particular, in trying to remove the   discrepancy issues in the gauge dependence of SIGWs, we find that the observable $\Omega_\text{GW}$ is actually gauge-independent in RD and MD.
We show that all the kernel functions lead to the same gauge independence.
Therefore, $\Omega_{\mathrm{GW}}$ should be identical in all the proposed gauge fixings. Hence the energy density $\Omega_\text{GW}$ of SIGWs converges as in the late time limit
$(x\gg1)$. Consequently, in principle, it indicates that the physical behavior of the observable $\Omega_{\mathrm{GW}}$ is the same in various gauges.  Moreover, SIGWs may explain the signal detected  by the North American Nanohertz Observatory for Gravitational Waves (NANOGrav) \cite{NANOGrav:2020gpb,Rezazadeh:2021clf}.

The article is organized as follows: In sect. \ref{sec.2}, we recapitulate the formalism for scalar-induced secondary GWs and the kernel  functions. Here,  we introduce the counter term that eliminates the extra scalar terms that cause discrepancies appearing in different gauges
during RD and MD. In sect. \ref{sec.2RD}, we explicitly investigate the kernel functions in various popular gauge choices during RD and MD, respectively. In sect. \ref{sec.3case}, in the sub-horizon modes,  we generally shed light on the physical behavior of the kernel functions to evaluate the observable energy density. In particular, we present the comparison between the transfer functions (which may cause discrepancies in some gauges) and kernel functions in different gauges.
Finally, we draw our discussion and concluding remarks in sect. \ref{conclus}.

\section{SIGWs in the RD and MD eras}
\label{sec.2}
In this section, we analyze the secondary tensor perturbations induced by first-order scalar perturbations in an expanding spacetime containing either pure RD or MD.
It is widely discussed in the literature that first-order tensor perturbations are gauge-independent. However, the secondary tensor perturbations may be gauge dependent \cite{Arroja:2009sh,Hwang:2017oxa,Tomikawa:2019tvi,Inomata:2019yww,Yuan:2019fwv,DeLuca:2019ufz,Lu:2020diy,Chang:2020iji,Nakamura:2011xy,Chang:2020mky},
while several gauge-invariant secondary tensor perturbations can be constructed in a particular gauge \cite{Bruni:1996im,Mollerach:2003nq,Bartolo:2004ty,Nakamura:2004rm,Nakamura:2006rk,Malik:2008im,Arroja:2009sh,Domenech:2017ems,Blaut:2019fxb,DeLuca:2019ufz,Chang:2020tji,Domenech:2020xin}.
Although SIGWs are usually discussed using the typically chosen Poisson gauge,  we need to study SIGWs in other popular gauges.

Armed with the results of SIGWs in the Poisson gauge \cite{Kohri:2018awv} 
we introduce a counter term in secondary tensor perturbations.
In the following sections, we analyze whether scalar-induced secondary GWs are gauge invariant or gauge dependent.
We investigate the energy density $\Omega_{\mathrm{GW}}$ of SIGWs in various gauges to answer the question.

In this subsection, we give the general formula for calculating SIGWs without specifying the background and the gauge.
To discuss SIGWs as the stochastic GW background, we consider the general perturbed metric around FLRW background as follows:
\begin{align}
g_{00}=&-a^2(1+2 \phi),\nonumber\\
g_{0i}=&\,2a^2 \partial_i B,\label{rdmetric}\\
g_{ij}=&\,a^2\delta_{ij}+a^2 \left(\frac{1}{2} h_{ij}^{\mathrm{TT}}-2\delta_{ij}\psi+2\partial_i \partial_j E\right)\nonumber,
\end{align}
where $a(\eta)$ is the scale factor of the universe. And the scalar perturbations $\phi$, $\psi$, $B$, and $E$ are of first order,
and the transverse traceless part $h_{ij}^{\mathrm{TT}}$ is the second-order tensor mode with $h^{\mathrm{TT}}_{ii}=0$ and $\partial_i h^\mathrm{TT}_{ij}=0$.

 To eliminate the fictitious terms in the secondary tensor perturbations by following refs. \cite{Nakamura:2020pre,Chang:2020tji}, we use gauge invariant variables by introduce
the counter term $\Xi_{kl}$ in the secondary tensor perturbation as:
\begin{equation}
   {\tilde h_{ij}}^{\mathrm{TT}}= h_{ij}^{\mathrm{TT}} + \mathcal{T}_{ij}^{kl} \Xi_{kl}\label{coneqn},
\end{equation}
where in the second term on the right-hand side of eq. \eqref{coneqn},   $\mathcal{T}_{ij}^{lm}=\Lambda_{i}^l\Lambda_{j}^m-\Lambda_{i}^l\Lambda_{ij}^{lm}/2$ represents the projection tensor that use to extract the transverse, trace-free part of a tensor 
and $\Xi_{kl}$ is defined as:
\begin{equation}
   \Xi_{kl} = -2\Big( 4E \partial_l \partial_k \phi +\partial^s E \partial_s \partial_l \partial_k E -(\partial_0 E- B)\partial_l \partial_k (\partial_0 E- B)\Big).\label{counterterm}
\end{equation}

Since $h_{ij}$ have been treated as gauge dependent with having divergence in the previous studies  \cite{DeLuca:2019ufz,Tomikawa:2019tvi, Lu:2020diy,Ali:2020sfw}. In contrast, here we are instead using ${\tilde h_{ij}}$ by the adding $\Xi_{kl}$ as a gauge-dependent counter term that, in principle, ensures the gauge independence of ${\tilde h_{ij}}$. Later, we have shown graphically that this counter term removes the divergence. The explicit expression of $\Xi_{kl}$, see e.g., refs. \cite{Malik:2008im,Nakamura:2011xy,Chang:2020tji} for details.

In the following, in the evaluations of SIGWs, first,  we briefly analyze the well-known results on $h_{ij}^{\mathrm{TT}}$. We consider that the generation of SIGWs starts long before the horizon reentry.
After  perturbing Einstein's equation $G_{\mu\nu}=8\uppi G T_{\mu\nu}$ up to the second-order, we get \cite{Chang:2020tji}
\begin{align}
{\tilde h_{ij}}^{\mathrm{TT}\prime\prime}+2\mathcal{H}{\tilde h_{ij}}^{\mathrm{TT}\prime}-\nabla^2{\tilde h_{ij}}^{\mathrm{TT}}=-4\mathcal{T}_{ij}^{lm}s_{lm}\label{theeq},
\end{align}
where  $s_{ij}$ is a source which is given by refs. \cite{Lu:2020diy,Ali:2020sfw}:
\begin{align}
\mathcal{T}_{ij}^{lm}s_{lm}=&\,{\partial_i}\psi{\partial_j}\psi+{\partial_i}\phi{\partial_j}\phi
-{\partial_i}{\partial_j}\sigma\left(\phi^\prime+\psi^\prime-\nabla^2\sigma\right)\notag\\
&+\left({\partial_i}\psi^\prime\sigma{\partial_j}+{\partial_j}\psi^\prime{\partial_i}\sigma\right)
-{\partial_i}{\partial_k}\sigma{\partial_j}{\partial_k}\sigma\notag\\
&+2{\partial_i}{\partial_j}\psi\left(\phi+\psi\right)
-8\uppi Ga^2({\rho_0}+{P_0}){\partial_i}\delta V{\partial_j}\delta V\notag\\
&-2{\partial_i}{\partial_j}\psi\nabla^2E
+2{\partial_i}{\partial_j}E\left(\psi^{\prime\prime}+2\mathcal{H}\psi^\prime-\nabla^2\psi\right)\notag\\
&-{\partial_i}{\partial_k}E^\prime {\partial_j}{\partial_k}E^\prime+{\partial_i}{\partial_k}{\partial_l}E{\partial_j}{\partial_k}{\partial_l}E\notag\\
&+2\left({\partial_j}{\partial_k}\psi{\partial_i}{\partial_k}E+{\partial_i}{\partial_k}\psi{\partial_i}{\partial_j}E\right)\notag\\
&-2\mathcal{H}({\partial_i}\psi{\partial_j}E^\prime+{\partial_j}\psi{\partial_i}E^\prime)
-\left({\partial_i}\psi^\prime {\partial_j}E^\prime+{\partial_j}\psi^\prime {\partial_i}E^\prime\right)\notag\\
&-\left({\partial_i}\psi{\partial_j}E^{\prime\prime}+{\partial_j}{\partial_i}\psi{\partial_j}E^{\prime\prime}\right)
+2{\partial_i}{\partial_j}E^\prime\psi^\prime\notag\\
&+{\partial_i}{\partial_j}{\partial_k}E{\partial_k}\left(E''+2\mathcal{H}E'-
\nabla^2E\right)\label{source11},
\end{align}
where $\sigma= E'-B$ is the shear potential,  the anisotropic stress tensor $\Pi_{ij}$ of the matter fluid is considered to be zero.

With different gauge choices, using $E=0$, the above source eq. \eqref{source11} can be reduced to the form given in refs. \cite{Gong:2019mui,DeLuca:2019ufz,Hwang:2017oxa} with negligible anisotropic stress.  Generally, we should use eq. \eqref{source11} instead. Especially here, we should include all those terms that contain $E$ in different gauges during RD and MD. In the Fourier space,  the tensor $h_{ij}^{\mathrm{TT}}$ can be expanded  with plus $\epsilon_{ij}^+$, and cross  $\epsilon_{ij}^\times$ polarization tensors as follows \cite{Lu:2020diy,Ali:2020sfw,Kohri:2018awv}:
\begin{equation}
\label{hijkeq1}
h_{ij}^{\mathrm{TT}}(\bm x,\eta)=\int\frac{\mathrm{d}^3k}{(2\uppi)^{3/2}}\epsilon^{\textrm{i}{\bm k}\cdot {\bm x}}[h^+_{\bm k}(\eta)\mathbf \epsilon^+_{ij}+{h}^\times_{\bm k}(\eta) \mathbf \epsilon^\times_{ij}].
\end{equation}
Next, we define the projection tensor for the source $s_{lm}(\bm x,\eta)$ in the Fourier space as:
\begin{equation}
\label{projeq}
\mathcal{T}_{ij}^{lm}s_{lm}=\int\frac{\mathrm{d}^3 k}{(2\uppi)^{3/2}}\epsilon^{\textrm{i} {\bm k} \cdot {\bm x}}[\mathbf \epsilon_{ij}^+ \mathbf \epsilon^{+lm}+\mathbf \epsilon_{ij}^\times \mathbf \epsilon^{\times lm}]s_{lm}(\bm k,\eta),
\end{equation}
we now find the solution to eq. \eqref{theeq} for  $\epsilon_{ij}^+$ as:
\begin{equation}
\label{hsolution}
h^+({\bm k},\eta)=4 \int\frac{\mathrm{d}^3p}{(2\uppi)^{3/2}} \mathbf \epsilon^{+ij}p_ip_j\zeta({\bm{p}})\zeta(\bm k-{\bm{p}})\frac{1}{k^2}I(u,v,x),
\end{equation}
where $x=k\eta$,
$u=p/k$, $v=\lvert{\bm k-\bm p}\rvert/k$, $\zeta(\bm{p})=\psi+\mathcal{H}\delta\rho/\rho_0^\prime$ is the primordial curvature perturbation. Here, one can assume equal contributions from the two polarized tensors in Fourier space. We use one polarization to evaluate its energy density and get the total energy density by doubling it. In the above expression \eqref{hsolution}, $I(u,v,x)$
is the kernel
function,  given by \cite{Kohri:2018awv,Inomata:2016rbd,Espinosa:2018eve,Lu:2019sti}
\begin{align}
\label{I_int}
I(u,v,x)=\int_0^x\mathrm{d}\tilde{x}\frac{a(\tilde{\eta})}{a(\eta)}kG_{k}(\eta,\tilde{\eta})f(u,v,\tilde x),
\end{align}
where $f(u,v,x)$ is associated with  $S_{\bm k}^+=\mathbf{e}^{+ij}s_{ij}(\bm k,\eta)$ as follows:
\begin{equation}
\label{sfrel1}
S_{\bm k}^+(\eta)=\int\frac{\mathrm{d}^3 p}{(2\uppi)^{3/2}}\zeta(\bm p)\zeta(\bm k-\bm p)\mathbf{e}^{+ij}p_i p_j f(u,v,x).
\end{equation}
From eqs. \eqref{theeq}-\eqref{sfrel1}, we derived these expressions in the semi-analytic way \cite{Kohri:2018awv,Lu:2020diy, Ali:2020sfw}. Furthermore, in the following, we derive the explicit expressions of the source function $f(u,v,x)$, which will be used in the subsequent sections to derive the kernel functions in seven different gauges. The source function $f(u,v,x)$ in eq. \eqref{sfrel1} can be symmetrized under the exchange  $u\leftrightarrow v$ for computational simplicity, as:
 \begin{equation}
     f(u,v,x)=\frac12(\Tilde{f}(u,v,x)+\tilde{f}(v,u,x)),
 \end{equation}
 where
 \begin{align}
         \Tilde{f}(u,v,x)=
         &\,T_\psi(ux)T_\psi(vx)-T_\phi(ux)T_\phi(vx)\notag\\
         &-\frac{v}{u}T_\sigma(ux)\left[T_\phi^*(vx)+T_\psi^*(vx)+T_\sigma(vx)\right]\notag\\
         &-2\frac{u}{v}T_\psi^*(ux)T_\sigma(vx)-\frac{1-u^2-v^2}{2uv}T_\sigma(ux)T_\sigma(vx)\nonumber\\
         &+2T_\psi(ux)T_\phi(vx)+\frac{2}{\mathcal{H}^2-\mathcal{H}}\notag\\
         &\times\left[kuT_\psi^*(ux)+\mathcal{H} T_\phi(ux)\right]\left[kvT_\psi^*(vx)+\mathcal{H}T_\phi(vx)\right]\notag\\
         &+2\frac{u^2}{v^2}T_E(vx)\left[T_\psi^{**}(ux)+\frac{2\mathcal{H}}{k  u}T_\psi^*(ux)+T_\psi(ux)\right]\nonumber\\
         &+ 2T_\psi(ux)T_E(vx)
         -\frac{1-u^2-v^2}{2uv}T_E^*(ux)T_E^*(vx)\notag\\
         &-\left(\frac{1-u^2-v^2}{2uv}\right)^2T_E(ux)T_E(vx)\notag\\
         &+4\frac{u}{v}T_\psi^*(ux)T_E^*(vx)+2T_\psi(ux)T_E^{**}(vx)\nonumber\\&
         +\frac{4\mathcal{H}}{k  v}T_\psi(ux)T_E^*(vx)
        -\frac{1-u^2-v^2}{2u^2}T_E(ux)\notag\\
         &\times\left[T_E^{**}(vx)+\frac{2\mathcal{H}}{k  v}T_E^*(vx)+T_E(vx)\right]\label{genrelf},
 \end{align}
and $T^*(y)=\mathrm{d}T(y)/\mathrm{d}y$.
The power spectrum of SIGWs then can be written as \cite{Lu:2020diy}:
\begin{align}
\label{PStensor}
\mathcal{P}_h(k,x)=&\,4\int_{0}^\infty\mathrm{d}u\int_{\lvert  1-v \rvert }^{1+u}\mathrm{d}v
\left[\frac{4u^2-(1+u^2-v^2)}{4uv}\right]^2\nonumber\\
         &\times I^{2}(u,v,x)\mathcal{P}_\zeta(uk)\mathcal{P}_\zeta(vk),
\end{align}
where $\mathcal{P}_\zeta$ is the primordial scalar power spectrum, and the GW  energy density $\rho_{\text{GW}}(\eta)=\int \text{d}\ln k \rho_{\text{GW}}(\eta, k)$ can be evaluated as \cite{Maggiore:1999vm}:
\begin{equation}
\rho_{\text{GW}}=\frac{{M_{pl}}^2}{16a^2} \left \langle \overline{h_{ij,k}h_{ij,k}} \right \rangle,  \label{rho_GW}
\end{equation}
where the over-line denotes the oscillation average.
In general, one can write the  fraction of the energy density of SIGWs as  \cite{Kohri:2018awv,Zhang:2020ptw}:
\begin{equation}\label{EGW0}
\Omega_{GW}\left(\eta,k\right)=\frac{\mathrm{d}\rho_{\mathrm{GW}}}{\rho_c\mathrm{d}\ln k}=\frac{1}{24}\left(\frac{k}{\mathcal{H\left(\eta\right)}}\right)^2\overline{\mathcal{P}_h\left(k,\eta\right)},
\end{equation}
where $\rho_c=3H^2/8\uppi G$ denote
the critical energy density of the Universe,
and
$\mathcal{P}_h$ can be defined as:
\begin{equation}
\left\langle h_{\bm k_1}^{s_1}(\eta)h_{\bm k_2}^{s_2}(\eta)\right\rangle=\frac{2\uppi^2}{k_1^3}\delta_{s_1s_2}\delta^3(\bm k_1+\bm k_2)\mathcal{P}_h(k_1,\eta),\ s_i=+,\times.
\end{equation}
Next, we analyze the secondary tensor perturbations under the various gauges during RD and MD. Here, we only consider the first-order scalars,  $\alpha$, and $\beta$. We do not consider the secondary coordinate transformation because the coordinate transformation of tensor modes does not depend on the transformation of the same order. Thus the secondary tensor perturbation
can be transformed with a counter term as \cite{Lu:2020diy, Ali:2020sfw, Zhang:2020ptw}:
\begin{equation}
\label{gaugetranf1}
{\tilde h_{ij}}^{\mathrm{TT}}\to h_{ij}^{\mathrm{TT}}+\chi_{ij}^{\mathrm{TT}}+\Xi_{ij}^{\mathrm{TT}},
\end{equation}
where
\begin{align}
\chi_{ij}^{\mathrm{TT}}(\bm x,\eta)&=\mathcal{T}_{ij}^{lm}\chi_{lm}\notag\\
         &=\int\frac{\mathrm{d}^3 k}{(2\uppi)^{3/2}}\textrm{e}^{{\rm i}\bm k\cdot \bm x}[\chi^+({\bm k}, \eta) \mathbf \epsilon_{ij}^++{\chi}^\times({\bm k}, \eta)\mathbf \epsilon_{ij}^\times],\label{xijtt}
\end{align}\vspace*{-4mm}
\begin{align}
\label{xijtt12}
\Xi_{ij}^{\mathrm{TT}}(\bm x,\eta)
=\,-\int\frac{\mathrm{d}^3 k}{(2\uppi)^{3/2}}\textrm{e}^{{\rm i}\bm k\cdot \bm x}[\Xi^+({\bm k}, \eta) \mathbf \epsilon_{ij}^++{\Xi}^\times({\bm k}, \eta)\mathbf \epsilon_{ij}^\times],
\end{align}\vspace*{-4mm}
\begin{align}
\chi^+({\bm k}, \eta)=&-\int\frac{\mathrm{d}^3p}{(2\uppi)^{3/2}} \mathbf \epsilon^{+ij}p_ip_j\bigg(4\alpha(\bm{p})\sigma(\bm k-\bm{p})+\frac{16}{\eta}\alpha(\bm{p})\notag\\
&\times[E(\bm k-\bm{p})+\beta(\bm k-\bm{p})]
\!+\!\bm{p}\cdot
(\bm k\!-\!\bm{p})\beta(\bm p)[4E(\bm k\!-\!\bm{p})\notag\\
&
+2\beta(\bm k-\bm{p})]-8\psi(\bm{p})\beta(\bm k-\bm{p})+2\alpha(\bm{p})\alpha(\bm k\!-\!\bm{p})\bigg),\notag\\
=&4\int\frac{\mathrm{d}^3p}{(2\uppi)^{3/2}}\mathbf \epsilon^{+ij}p_ip_j \zeta(\bm{p})\zeta(\bm k-\bm{p})\frac{1}{k^2}I_\chi(u,v,x)\label{chi_fourier},
\end{align}
and
\begin{align}
\Xi^+({\bm k}, \eta)
=&\,-\int\frac{\mathrm{d}^3p}{(2\uppi)^{3/2}}\mathbf \epsilon^{+ij}p_ip_j {3 (1 + w)}/{(5 + 3 w)}\notag\\
	&\times\zeta(\bm{p})\zeta(\bm k-\bm{p})\frac{1}{k^2}I_\Xi(u,v,x),\label{56_fourier}
\end{align}
where  $I_{\chi} \left( u,v,x\right)$ takes the form of
 \begin{align}
   I_\chi(u,v,x)=& - \frac{1}{9uv}\Bigg[\vphantom{\frac{u^2}{v^2}}2T_\alpha(ux)T_\sigma(vx)  \notag\\
&    +2T_\alpha(vx)T_\sigma(ux)+2T_\alpha(ux)T_\alpha(vx)\notag\\
& -4\left(\frac{u}{v}T_\psi(ux)T_\beta(vx)+\frac{v}{u}T_\psi(vx)T_\beta(ux)\right)\notag\\
   &+\frac{1-u^2-v^2}{uv}[T_\beta(ux)T_E(vx)\notag\\
& +T_\beta(vx)T_E(ux)
   +T_\beta(ux)T_\beta(vx)]\notag\\
& +4\frac{\mathcal{H}}{k}
   \left(\frac1{v} T_\alpha(ux)T_E(vx)+\frac1u T_E(ux)T_\alpha(vx)\right.\notag\\
   &\left.\left.+\frac1{v} T_\alpha(ux)T_\beta(vx)+\frac1u T_\beta(ux)T_\alpha(vx)\right)\right], \label{Ichi}
 \end{align}
while the kernel function $I_{\Xi} \left( u,v,x\right)$ takes the form of
 \begin{align}
	I_{\Xi} \left( u, v, x \right)   =& -2\left(- \frac{2 }{v^2}
	T_E \left( vx \right) T_{\psi} (ux) - \frac{2}{u^2} T_{\psi} \left( vx \right) T_E (ux)\right.\notag\\
&
 \left.+ \frac{2}{uv} T_B \left(vx \right) T_B (ux)+  \left( \frac{1}{v^2} T_E'
	\left( vx \right) - \frac{1}{v} T_B \left(vx \right) \right)\right.\notag\\&
\times\left. \left( \frac{1}{u^2} T_E' (ux) -
	\frac{1}{u} T_B (ux) \right)\right.\notag\\ &
\left.+  \left( \frac{u \mathcal{H}}{v (k)^2} \right)  T_E \left( vx \right) T_E (ux)\right). \label{Xi}
 \end{align}
Here we symmetrized the kernel function $I_\chi(u,v,x)$ under $u\leftrightarrow v$.
It is worth mentioning that the transformed secondary tensor perturbations have expressions in the form of the first-order scalar coordinate transformation. With the gauge transformation  \eqref{gaugetranf1} and the result for SIGWs in the Poisson gauge,  it is easy to accomplish the (semi)analytic derivation of the SIGWs in any chosen gauge without conducting the complicated calculations in that gauge.
Combining eqs. \eqref{hijkeq1}, \eqref{hsolution}, \eqref{gaugetranf1}, \eqref{xijtt}, \eqref{chi_fourier}, and \eqref{Xi} one obtains the gauge transformation of SIGWs as follows:
\begin{align}\label{hchi}
  &\tilde{h}^+_{\bm k}\rightarrow h^+_{\bm k}+\chi^+_{\bm k}+\Xi^+_{\bm k}\notag\\
  &=4\int\frac{\mathrm{d}^3p}{(2\uppi)^{3/2}}\mathbf e^{+ij}(\bm k)p_ip_j\zeta(\bm p)\zeta(\bm k-\bm p)\notag\\
  &\quad\times\frac{1}{k^2}\left[I(u,v,x)+I_\chi(u,v,x)+I_\Xi(u,v,x)\right],
\end{align}
and for the perturbations, one can use the transfer functions $T(x)$ as follows:
\begin{align}
\label{defs}
\alpha(\bm k,x)&=\frac{3(1+w)}{5+3w}\zeta(\bm k)\frac{1}{k}T_\alpha(x),\\
\beta(\bm k,x)&=\frac{3(1+w)}{5+3w}\zeta(\bm k)\frac{1}{k^2}T_\beta(x),\\
\sigma(\bm k,x)&=\frac{3(1+w)}{5+3w}\zeta(\bm k)\frac{1}{k}T_\sigma(x),\\
E(\bm k,x)&=\frac{3(1+w)}{5+3w}\zeta(\bm k)\frac{1}{k^2}T_E(x),\\
B(\bm k,x)&=\frac{3(1+w)}{5+3w}\zeta(\bm k)\frac{1}{k}T_B(x),\\
\psi(\bm k,x)&=\frac{3(1+w)}{5+3w}\zeta(\bm k)T_\psi(x),\\
\phi(\bm k,x)&=\frac{3(1+w)}{5+3w}\zeta(\bm k)T_\phi(x).
\end{align}
 The above gauge transformation  \eqref{hchi} is our main result for studying the gauge transformation  of SIGWs in general. From the expression \eqref{hchi}, one can either transform the solution or show how the power spectrum of SIGWs can be transformed  under the gauge transformation. For instance, with the solution in the Poisson gauge,
 one can get the solution in
any gauge, according to the following transformation:
\begin{align}
I_{\tilde h}(u,v,x) \to I_{\tilde h}(u,v,x)+I_\chi(u,v,x)+I_\Xi (u,v,x).\label{INc}
\end{align}
Here, during RD, we have 
\begin{align}
&I_{\text{RD},\,{\chi}}(u,v,x)= - \frac{1}{9uv}\left[\vphantom{\frac{u^2}{v^2}}
-4\left(\frac{u}{v}T_\mathrm{P}(ux)T_\beta(vx)
+\frac{v}{u}T_\mathrm{P}(vx)T_\beta(ux)\right)\right.\nonumber\\
&\quad \left.
+2T_\alpha(ux)T_\alpha(vx)+
\frac{4}{k\eta}
\left(\frac1v T_\alpha(ux)T_\beta(vx)+\frac1u T_\beta(ux)T_\alpha(vx)\right)
\right.\nonumber\\
&\quad \left.+\frac{1-u^2-v^2}{uv}T_\beta(ux)T_\beta(vx) \right]\label{ichi_newton},
\end{align}\vspace*{-4mm}
\begin{align}
&I_{\text{RD},\, \tilde h}(u,v,x)= - \frac{1}{9uv}\left[\vphantom{\frac{u^2}{v^2}}
-4\left(\frac{u}{v}T_\mathrm{P}(ux)T_\beta(vx)
+\frac{v}{u}T_\mathrm{P}(vx)T_\beta(ux)\right)\right.\nonumber\\
&\quad \left.
+2T_\alpha(ux)T_\alpha(vx)+
\frac{4}{k\eta}
\left(\frac1v T_\alpha(ux)T_\beta(vx)+\frac1u T_\beta(ux)T_\alpha(vx)\right)
\right.\nonumber\\
&\quad \left.+\frac{1-u^2-v^2}{uv}T_\beta(ux)T_\beta(vx) +18 T_B \left(vx \right) T_B (ux)\right]\label{ichi_Xi},
\end{align}
and for MD is
\begin{align}
   I_{\text{MD},\,\chi}(u,v,x)=& - \frac{9}{100uv}\left[\vphantom{\frac{u^2}{v^2}}
   2T_\alpha(ux)T_\alpha(vx)+ u^2v^2T_\beta(ux)T_\beta(vx) \right.\notag\\
   &-4\left(\frac{u^2+v^2}{uv}\left(T_\mathrm{P}(ux)T_\beta(vx)
   +\frac{v}{u}T_\mathrm{P}(vx)T_\beta(ux)\right)\right)\notag\\
   &+\frac{8}{x}
   \left(\frac1{v} T_\alpha(ux)T_\beta(vx)+\frac1u T_\beta(ux)T_\alpha(vx)\right)\notag\\
&   \left.-\frac{u^2+v^2}{uv}T_\beta(ux)T_\beta(vx)\right],
\label{ichi_newtoncase}
\end{align}\vspace*{-4mm}
\begin{align}
   I_{\text{MD},\,\tilde{h}}(u,v,x)=& - \frac{9}{100uv}\left[\vphantom{\frac{u^2}{v^2}}
   2T_\alpha(ux)T_\alpha(vx)+ u^2v^2T_\beta(ux)T_\beta(vx) \right.\notag\\
   &-4\left(\frac{u^2+v^2}{uv}\left(T_\mathrm{P}(ux)T_\beta(vx)
   +\frac{v}{u}T_\mathrm{P}(vx)T_\beta(ux)\right)\right)\notag\\
   &+\frac{8}{x}
   \left(\frac1{v} T_\alpha(ux)T_\beta(vx)+\frac1u T_\beta(ux)T_\alpha(vx)\right)\notag\\
&   \left.-\frac{u^2+v^2}{uv}T_\beta(ux)T_\beta(vx) +\frac{200}{9} T_B \left(vx \right) T_B (ux)\right].
\label{ichi_newtoncaseXi}
\end{align}
The above expressions \eqref{ichi_newton} and \eqref{ichi_newtoncase} can be obtained by the use of the  transfer functions $T_\sigma=T_E=0$ and $T_\psi=T_\text{P}$
 into eq. \eqref{Ichi}, in the Poisson  gauge. Besides,  gauge transformation   from the Poisson
gauge to the other gauges gives the transfer functions $T_\alpha$ and $T_\beta$, respectively. In the following, we explicitly use eq. \eqref{INc} to evaluate the kernel function in any given gauge.
\section{Results of the Kernel Functions during RD and MD}
\label{sec.2RD}
This section presents the results of the kernel functions by evaluating the transfer functions of the metric perturbations in seven different gauges. These transfer functions describe the evolution of the density perturbations on subhorizon scales.
 One can see the gauge (in)dependence by evaluating the kernel function $I(u,v,x)$. 
\subsection{Poisson gauge}
\label{ngrd}
In this subsection,  we consider the standard  Poisson  gauge with $B=E=0$. Here we use the Bardeen's potentials  $\phi_\mathrm{P}=\psi_\mathrm{P}=\Phi=\Psi$ \cite{Bardeen:1980kt}. 
In the RD universe, we have $\Phi=\Psi=2\zeta/3$ on superhorizon scales. Conversely, the counter perturbation $\Xi_{kl}$ vanishes in the Poisson gauge i.e., $\Xi_{kl}=0$ \cite{Chang:2020tji}.  Following ref. \cite{Lu:2020diy} one can calculate the analytical expression of $I(u,v,x)$
explicitly at the late time, $x \gg 1$  as follows:
\begin{align}
&I_{\text{RD,P}}(u,v,x)\notag\\
&=
-\frac{3}{u^3v^3x^4}\left(
(ux)^3v\sin x+u(vx)^3\sin x-3uvx^3\sin x\right)\notag\\&
\quad-\frac{3}{u^3v^3x^4}\bigg(6uvx^2\cos\frac{ux}{\sqrt{3}}\cos\frac{vx}{\sqrt{3}}+6\sqrt{3}ux\cos\frac{ux}{\sqrt{3}}\sin\frac{vx}{\sqrt{3}}\nonumber\\
&\quad\left.-18\sin\frac{ux}{\sqrt{3}}\sin\frac{vx}{\sqrt{3}}+(u^2+v^2-3)x^2\sin\frac{ux}{\sqrt{3}}\sin\frac{vx}{\sqrt{3}}
\right)\notag\\&
\quad\times\left[
\left(
\mathrm{Ci}\left[\left(1\!+\!\frac{u-v}{\sqrt{3}}\right)x\right]\!+\!
\mathrm{Ci}\left[\left(1\!+\!\frac{v-u}{\sqrt{3}}\right)x\right]\!-\!
\mathrm{Ci}\left[\left(1\!+\!\frac{u+v}{\sqrt{3}}\right)x\right]\right.\right.\notag\\
&\quad\left.
-\mathrm{Ci}\left[\left|1-\frac{u+v}{\sqrt{3}}\right|x\right]
+\ln\left[
\left|\frac{3-(u+v)^2}{3-(u-v)^2}\right|
\right]\right)\sin x\notag\\&
\quad+\left(
-\mathrm{Si}\left[\left(1+\frac{u-v}{\sqrt{3}}\right)x\right]\right.
-
\mathrm{Si}\left[\left(1+\frac{v-u}{\sqrt{3}}\right)x\right]\notag\\&
\quad+
\mathrm{Si}\left[\left(1-\frac{u+v}{\sqrt{3}}\right)x\right]
+\left.\left.
\mathrm{Si}\left[\left(1+\frac{u+v}{\sqrt{3}}\right)x\right]
\right)\cos x
\right]
\label{I_P}.
\end{align}
The subscript ``$\mathrm{P}$" indicates the evaluation in the Poisson gauge.
The evolution of $I^2_\text{P}(u,v,x)$ with $u=v=1$ and $u=v=0.1$ is shown in Figures \ref{fig:divergentRD} and \ref{fig:convergentRD}, respectively. It is to be noted  that $I_{\mathrm{P}}(u,v,x\rightarrow\infty)\propto x^{-1}$,
and $\Omega_{\mathrm{GW}}(k,x\rightarrow\infty)$ is a constant. It means that SIGWs appear as free radiation deep inside the horizon.

In MD $w=0$, the Bardeen's potentials are   $\Phi=\Psi=3\zeta/5$.  In this gauge,  the kernel function can be expressed explicitly as:
\begin{equation}\label{I_Nmd}
   I_{\mathrm{MD,P}}(u,v,x)=\frac{18(x\cos x -\sin x)}{5x^3}+\frac{6}{5}.
\end{equation}

As $I_{\mathrm{P}}(u,v,x\rightarrow\infty)=6/5$,  eq. \eqref{PStensor} shows that the primordial power spectrum
$\mathcal{P}_h$ is a constant at $x\gg1$, and the energy density $\Omega_{\mathrm{GW}}$  is proportional to  $x^2$. Hence
$\Omega_{\mathrm{GW}}(k,x\rightarrow\infty)\propto a$ if we use eq. \eqref{rho_GW}.
According to eq. \eqref{hsolution}, it can be seen that the constant term $6/5$ in eq. \eqref{I_Nmd}
contributes a constant to $h_{\bm k}$. Consequently, the contribution to
$h_{\bm k}'$ and the energy density approaches zero. It means that we should use the definition \eqref{EGW0} to determine $\Omega_{\text{GW}}$. Otherwise, the constant $6/5$ will be mistakenly calculated if we use eq. \eqref{rho_GW}.
Accordingly, the constant $6/5$ in eq. \eqref{I_Nmd}
does not provide any contribution to the energy density $\Omega_{\mathrm{GW}}$.
Thus, the constant in eq. \eqref{I_Nmd} does not represent a wave solution, and GWs come from those terms that represent the oscillations as $\sin x$ and $\cos x$.
After barring the constant factor $6/5$, one can find  $I_{\mathrm{P}}(x\rightarrow \infty)\propto \cos x/x^2=\cos x/a$ that leads to $\Omega_{\text{GW}}\propto a^{-1}$ and $\rho_{\text{GW}}\propto a^{-4}$,
which behaves, as one would expect, as radiation in the MD era.
However, only the terms with the oscillations as $\sin x$ and $\cos x$ provide evidence for SIGWs.

Since the energy density $\Omega_{\text{GW}}$ of SIGWs is uniquely determined by the investigation of the kernel functions in different gauge choices, the energy density spectrum evaluated in the gauge-independent framework takes the same form as those examined in the Poisson gauge during RD and\linebreak MD.
\subsection{TT gauge}
The TT gauge can be defined as $\phi=B=0$.
Since the kernel function $I(u,v,x)$ depends on the source function $f(u,v,x)$ linearly presented in eq. \eqref{I_int}. In this gauge, we can find the transfer function as follows:
\end{multicols}
\noindent\rule{85.5mm}{0.4pt}\rule{0.4pt}{2mm}
\vspace*{2mm}
\begin{align}
&I_{\mathrm{RD,\,TT}}(u,v,x)
=-\frac{3}{u^3v^3x^4}\left(
(ux)^3v\sin x+u(vx)^3\sin x-3uvx^3\sin x\right)
\!+\!\frac{3(u^2+v^2-3)^2}{4u^3v^3x}\times\ln\left[
\Big\lvert \frac{3-(u+v)^2}{3-(u-v)^2}\Big\rvert
\right] \sin x \!+\!\frac{3(u^2+v^2-3)^2}{4u^3v^3x}\notag\\[6pt]
&\times\left(-6uvx^2\cos\frac{ux}{\sqrt{3}}\cos\frac{vx}{\sqrt{3}}+6\sqrt{3}ux\cos\frac{ux}{\sqrt{3}}\sin\frac{vx}{\sqrt{3}}
+6\sqrt{3}vx\sin\frac{ux}{\sqrt{3}}\cos\frac{vx}{\sqrt{3}}-18\sin\frac{ux}{\sqrt{3}}\sin\frac{vx}{\sqrt{3}}+(u^2+v^2-3)x^2\right.\nonumber\\[6pt]
&\times\left.\sin\frac{ux}{\sqrt{3}}\sin\frac{vx}{\sqrt{3}}
\right)-\frac{9}{u^2v^2x^2}\left(
(1-u^2-v^2)x^2\left[
\mathrm{Ci}\left(\frac{ux}{\sqrt{3}}\right)+\mathcal{C}-\ln\frac{ux}{\sqrt{3}}-\frac{\sin(ux/\sqrt{3})}{ux/\sqrt{3}}
\right]\right.
 \times\left[\mathrm{Ci}\left(\frac{vx}{\sqrt{3}}\right)\!+\!\mathcal{C}
\!-\!\ln\frac{vx}{\sqrt{3}}\!-\!\frac{\sin(vx/\sqrt{3})}{vx/\sqrt{3}}\right]\nonumber\\[6pt]
&+2\left[\frac{\sin(ux/\sqrt{3})}{ux/\sqrt{3}}-1\right]
\left[\frac{\sin(vx/\sqrt{3})}{vx/\sqrt{3}}-1\right]
+4\left[-\mathrm{Ci}\left(\frac{ux}{\sqrt{3}}\right)
\!-\!\mathcal{C}\!+\!\ln\frac{ux}{\sqrt{3}}\!+\!\frac{\sin(ux/\sqrt{3})}{ux/\sqrt{3}}\right]
\left[1-\cos\frac{vx}{\sqrt{3}}\right]
\nonumber\\[6pt]
&\left.+4\left[-\mathrm{Ci}\left(\frac{vx}{\sqrt{3}}\right)-\mathcal{C}+\ln\frac{vx}{\sqrt{3}}
+\frac{\sin(vx/\sqrt{3})}{vx/\sqrt{3}}\right]
\left[1-\cos\frac{ux}{\sqrt{3}}\right]
\right)
\times\left[
\left(
\mathrm{Ci}\left[\left(1\!+\!\frac{u-v}{\sqrt{3}}\right)x\right]+
\mathrm{Ci}\left[\left(1\!+\!\frac{v-u}{\sqrt{3}}\right)x\right]-
\mathrm{Ci}\left[\left(1\!+\!\frac{u+v}{\sqrt{3}}\right)x\right]\right.\right.\nonumber\\[6pt]
&\left.
-\mathrm{Ci}\left[\left\lvert 1-\frac{u+v}{\sqrt{3}}\right\rvert x\right]
\right)+\left(
-\mathrm{Si}\left[\left(1+\frac{u-v}{\sqrt{3}}\right)x\right]-
\mathrm{Si}\left[\left(1+\frac{v-u}{\sqrt{3}}\right)x\right]\right.
\left.\left.
+\mathrm{Si}\left[\left(1-\frac{u+v}{\sqrt{3}}\right)x\right]+
\mathrm{Si}\left[\left(1+\frac{u+v}{\sqrt{3}}\right)x\right]
\right)\cos x
\right]
\label{TT_check}.
\end{align}
\begin{multicols}{2}
In this gauge,  as mentioned above, the expression \eqref{TT_check} is obtained by using values of the transfer functions.
 In fact, as we show in Figure \ref{fig:divergentRD} in this work, they lead to a discrepancy between the Poisson gauge and the TT gauge. In the refs. \cite{Hwang:2017oxa, DeLuca:2019ufz, Lu:2020diy, Tomikawa:2019tvi, Ali:2020sfw}, the secondary tensor perturbations generated by the quadratic combination of a linear scalar-type cosmological perturbation are widely\linebreak investigated.

Nevertheless, most previous studies are based on a Poisson gauge without proper explanation.
 There in the previous studies, it is shown that the secondary induced tensor perturbations are generically gauge dependent.
In addition,  it is also presented over there that the result of the kernel function of the Poisson gauge is different from other gauges. Similarly, in this work, we find that the kernel in the TT gauge is different from the Poisson gauge. And there are also some pure gauge modes \cite{Lu:2020diy, Ali:2020sfw}. These different results with pure gauge modes cause a divergence. Here,  we try to fix the discrepancy by using a counter term given in the aforementioned expression \eqref{hchi}, and we show that being second order in perturbation, such induced tensor perturbations are generically gauge independent in contrast to refs. \cite{Hwang:2017oxa, DeLuca:2019ufz, Lu:2020diy, Tomikawa:2019tvi, Ali:2020sfw}, as shown in Figure  \ref{fig:convergentRD}.

Therefore, by using the transformation with a counter term, the kernel function of the gauge independent SIGWs in eq. \eqref{INc} is obtained  to be
\end{multicols}
\noindent\rule{85.5mm}{0.4pt}\rule{0.4pt}{2mm}
\begin{align}
&I_{\text{RD},\,\tilde{h},\, \text{TT}}(u,v,x)
=-\frac{3}{u^3v^3x^4}\left(
(ux)^3v\sin x+u(vx)^3\sin x-3uvx^3\sin x\right)
-\frac{3}{u^3v^3x^4}\Bigg(6uvx^2\cos\frac{ux}{\sqrt{3}}\cos\frac{vx}{\sqrt{3}}\nonumber\\
&\left.+6\sqrt{3}ux\cos\frac{ux}{\sqrt{3}}\sin\frac{vx}{\sqrt{3}}-18\sin\frac{ux}{\sqrt{3}}\sin\frac{vx}{\sqrt{3}}+(u^2+v^2-3)x^2\sin\frac{ux}{\sqrt{3}}\sin\frac{vx}{\sqrt{3}}
\right)
\times\left[
\left(
\mathrm{Ci}\left[\left(1+\frac{u-v}{\sqrt{3}}\right)x\right]+
\mathrm{Ci}\left[\left(1+\frac{v-u}{\sqrt{3}}\right)x\right]\right.\right.\notag\\
&\left.-
\mathrm{Ci}\left[\left(1+\frac{u+v}{\sqrt{3}}\right)x\right]
-\mathrm{Ci}\left[\left|1-\frac{u+v}{\sqrt{3}}\right|x\right]
+\ln\left[
\left|\frac{3-(u+v)^2}{3-(u-v)^2}\right|
\right]\right)\sin x+\left(
-\mathrm{Si}\left[\left(1+\frac{u-v}{\sqrt{3}}\right)x\right]-
\mathrm{Si}\left[\left(1+\frac{v-u}{\sqrt{3}}\right)x\right]\right.\notag\\
&
+
\mathrm{Si}\left[\left(1-\frac{u+v}{\sqrt{3}}\right)x\right]\left.\left.+
\mathrm{Si}\left[\left(1+\frac{u+v}{\sqrt{3}}\right)x\right]
\right)\cos x
\right]
\label{I_PTT}.
\end{align}
\noindent\hspace{92.5mm}\vrule width0.4pt height0.4pt depth 2mm\rule{85.5mm}{0.4pt}
\begin{multicols}{2}
 The counter-term in the TT gauge is non-trivial.
Besides, as we can see, the terms in the first two lines of the expression \eqref{TT_check} are the same with eq. \eqref{I_PTT}. Suppose we had the terms that are freely propagating tensor perturbations and only represented the free GWs. In that case, these free oscillating terms contribute to the energy density $\Omega_{\text{GW}}$ of SIGWs \cite{Inomata:2019yww}.  The evolution of the divergent kernel function $I^{2}_\text{RD,\,TT}(u,v,x)$ with $u=v=1$ and $u=v=0.1$ is shown in Figure \ref{fig:divergentRD} and for the
gauge independent kernel $I^2_{\tilde{h},\,\text{RD,\,TT}}(u,v,x)$ of the energy density of SIGWs is shown in Figure \ref{fig:convergentRD}, respectively.

 We finally obtain the kernel function in the MD universe from eq. \eqref{ichi_newtoncase} as follows:
\begin{align}
\label{TTkernelmd}
I_{\text{MD,\, TT}}(u,v,x)
=&\frac{18(x\cos x -\sin x)}{5x^3}\notag\\&+\frac{2400x^3+5x^5\Bigl(
-88+(-1+u^2+v^2)x^2\Bigr)}{2000x^3}.
\end{align}
While during MD, by using eq. \eqref{INc}, we can evaluate the independent  kernel function in the TT gauge as:
\begin{equation}
\label{TTkernelmdct}
I_{{\text{MD},\,\tilde{h},\, \text{TT}}}(u,v,x)
=\frac{18(x\cos x -\sin x)+30x^{3}}{5x^3}.
\end{equation}
It is to be noted that in the above expression \eqref{TTkernelmd}, the first term is the same as appearing in eq. \eqref{TTkernelmdct}. The only oscillating terms $\sin{x}$ or $\cos{x}$ contribute to SIGWs and show the physical behavior of the energy density of SIGWs, and the rest are fictitious terms.
The evolution of the kernel function \eqref{TTkernelmd}  is presented in Figures \ref{fig:divergent.MD} and \ref{fig:convergent.MD}. And the evolution for the gauge independent kernel function \eqref{TTkernelmdct} is shown in  Figures \ref{fig:divergent.MDbarringfactor} and \ref{fig:convergent.MDafterbarringfactor}, respectively.
\subsection{Comoving orthogonal gauge}
 Let us consider the comoving orthogonal gauge defined by $\delta V=B=0$. We use the results obtained from the background equations.
 After some algebraic calculations,
 we obtain the $I(u,v,x)$ in the comoving orthogonal gauge as:
\end{multicols}
\noindent\rule{85.5mm}{0.4pt}\rule{0.4pt}{2mm}
\begin{align}
&I_{\text{RD,\,CO}}\left(u,v,x\right)=
-\frac{3}{u^3v^3x^4}\left(
(ux)^3v\sin x+u(vx)^3\sin x-3uvx^3\sin x\right)
+\frac{3(u^2+v^2-3)^2}{4u^3v^3x}\times\ln\left[
\Big\lvert \frac{3-(u+v)^2}{3-(u-v)^2}\Big\rvert
\right] \sin x \notag\\&+\frac{3}{4u^3v^3x^4}\left[
3\mathcal{C}^2uv(u^2+v^2-1)x^4-2\sqrt{3}\mathcal{C}v(5u^2+3v^2-3)x^3\sin\frac{ux}{\sqrt{3}}-2\sqrt{3}\mathcal{C}u(3u^2+5v^2-3)x^3\sin\frac{vx}{\sqrt{3}}\right.\nonumber\\
&
-2[36-18(2u^2+2v^2-1)x^2+u^2v^2x^4]\sin\frac{ux}{\sqrt{3}}\sin\frac{vx}{\sqrt{3}}+3uvx^2[-8+(u^2+v^2-1)x^2]\cos\frac{ux}{\sqrt{3}}\cos\frac{vx}{\sqrt{3}}
\nonumber\\
&+vx\cos\frac{vx}{\sqrt{3}}\left(3\mathcal{C}u(u^2+v^2-1)x^3
-2\sqrt{3}[-12+(7u^2+3v^2-3)x^2]\sin\frac{ux}{\sqrt{3}}\right)\nonumber\\
&\left.+ux\cos\frac{ux}{\sqrt{3}}
\left(
3\mathcal{C}v(u^2+v^2-1)x^3
-2\sqrt{3}[-12+(3u^2+7v^2-3)x^2]\sin\frac{vx}{\sqrt{3}}\right)
\right]\notag\\
&+\frac{3}{u^3v^3x^4}\left(-6uvx^2\cos\frac{ux}{\sqrt{3}}\cos\frac{vx}{\sqrt{3}}+6\sqrt{3}ux\cos\frac{ux}{\sqrt{3}}\sin\frac{vx}{\sqrt{3}}
+6\sqrt{3}vx\sin\frac{ux}{\sqrt{3}}\cos\frac{vx}{\sqrt{3}}-18\sin\frac{ux}{\sqrt{3}}\sin\frac{vx}{\sqrt{3}}\right.\nonumber\\
&\left.+(u^2+v^2-3)x^2\sin\frac{ux}{\sqrt{3}}\sin\frac{vx}{\sqrt{3}}
\right)+\frac{3(u^2+v^2-3)^2}{4u^3v^3x}
\times\left[
\left(
\mathrm{Ci}\left[\left(1+\frac{u-v}{\sqrt{3}}\right)x\right]+
\mathrm{Ci}\left[\left(1+\frac{v-u}{\sqrt{3}}\right)x\right]-
\mathrm{Ci}\left[\left(1+\frac{u+v}{\sqrt{3}}\right)x\right]\right.\right.\nonumber\\
&\left.
-\mathrm{Ci}\left[\left\lvert 1-\frac{u+v}{\sqrt{3}}\right\rvert x\right]
\right)\sin x+\left(
-\mathrm{Si}\left[\left(1+\frac{u-v}{\sqrt{3}}\right)x\right]-
\mathrm{Si}\left[\left(1+\frac{v-u}{\sqrt{3}}\right)x\right]\right.
\left.\left.
+
\mathrm{Si}\left[\left(1-\frac{u+v}{\sqrt{3}}\right)x\right]+
\mathrm{Si}\left[\left(1+\frac{u+v}{\sqrt{3}}\right)x\right]
\right)\cos x
\right].\label{ICOM}
\end{align}
\noindent\hspace{92.5mm}\vrule width0.4pt height0.4pt depth 2mm\rule{85.5mm}{0.4pt}
\begin{multicols}{2}
 In the above expression \eqref{ICOM}, there exist some extra terms that do not contribute to SIGWs and cause a divergence, as it is shown in Figure \ref{fig:divergentRD}.
 We try to present a resolution by using a counter term in the aforementioned expression \eqref{hchi}. We show that being second order in perturbation, such induced tensor perturbations are generically gauge independent in contrast to refs. \cite{Hwang:2017oxa, Lu:2020diy, Tomikawa:2019tvi, Ali:2020sfw}, as shown in Figure \ref{fig:convergentRD}.

Therefore, by using the transformation of a counter term, the kernel function of the gauge independent SIGWs in eq. \eqref{INc} is obtained  to be
\end{multicols}
\noindent\rule{85.5mm}{0.4pt}\rule{0.4pt}{2mm}
\begin{align}
&I_{\text{RD},\,\tilde{h},\, \text{CO}}(u,v,x)
=-\frac{3}{u^3v^3x^4}\left(
(ux)^3v\sin x+u(vx)^3\sin x-3uvx^3\sin x\right)
-\frac{3}{u^3v^3x^4}\Bigg(6uvx^2\cos\frac{ux}{\sqrt{3}}\cos\frac{vx}{\sqrt{3}}\nonumber\\
&\left.+6\sqrt{3}ux\cos\frac{ux}{\sqrt{3}}\sin\frac{vx}{\sqrt{3}}-18\sin\frac{ux}{\sqrt{3}}\sin\frac{vx}{\sqrt{3}}+(u^2+v^2-3)x^2\sin\frac{ux}{\sqrt{3}}\sin\frac{vx}{\sqrt{3}}
\right)
\times\left[
\left(
\mathrm{Ci}\left[\left(1+\frac{u-v}{\sqrt{3}}\right)x\right]+
\mathrm{Ci}\left[\left(1+\frac{v-u}{\sqrt{3}}\right)x\right]\right.\right.\notag\\[6pt]
&\left.-
\mathrm{Ci}\left[\left(1+\frac{u+v}{\sqrt{3}}\right)x\right]
-\mathrm{Ci}\left[\left|1-\frac{u+v}{\sqrt{3}}\right|x\right]
+\ln\left[
\left|\frac{3-(u+v)^2}{3-(u-v)^2}\right|
\right]\right)\sin x+\left(
-\mathrm{Si}\left[\left(1+\frac{u-v}{\sqrt{3}}\right)x\right]-
\mathrm{Si}\left[\left(1+\frac{v-u}{\sqrt{3}}\right)x\right]\right.\notag\\[6pt]
&+
\mathrm{Si}\left[\left(1-\frac{u+v}{\sqrt{3}}\right)x\right]\left.\left.+
\mathrm{Si}\left[\left(1+\frac{u+v}{\sqrt{3}}\right)x\right]
\right)\cos x
\right]
\label{I_Pcomo112}.
\end{align}
\noindent\hspace{92.5mm}\vrule width0.4pt height0.4pt depth 2mm\rule{85.5mm}{0.4pt}
\begin{multicols}{2}
 The counter-term in the comoving orthogonal gauge is non-trivial.
Besides, as we can see, the terms in the first two lines of the expression \eqref{ICOM} are the same with eq. \eqref{I_Pcomo112}. Suppose we had the terms that are freely propagating tensor perturbations and only represented the free GWs. In that case, these free oscillating terms contribute to the energy density $\Omega_{\text{GW}}$ of SIGWs \cite{Inomata:2019yww}.  The evolution of the divergent kernel function $I^2_\text{RD,\,CO}(u,v,x)$ with $u=v=1$ and $u=v=0.1$ is shown in Figure \ref{fig:divergentRD} and for the
gauge independent kernel $I^2_{\tilde{h},\,\text{RD,\,CO}}(u,v,x)$ of the energy density of SIGWs is shown in Figure \ref{fig:convergentRD}, respectively.

In the MD universe, to calculate the result of the kernel function, we use eq. \eqref{ichi_newtoncase} and get
\begin{align}
I_{\text{MD,\, CO}}&(u,v,x)
=\frac{18(x\cos x -\sin x)}{5x^3}\notag\\&\qquad+\frac{2400x^3+5x^5\Bigl(
-88+(-1+u^2+v^2)x^2\Bigr)}{2000x^3}.\label{comocont}
\end{align}
While during MD, by using eq. \eqref{INc}, we can evaluate the independent  kernel function in the comoving orthogonal gauge as:
\vspace*{2mm}
\begin{equation}
\label{TTkernelmdctcom112}
I_{\text{MD},\,\tilde{h},\, \text{CO}}(u,v,x)
=\frac{18(x\cos x -\sin x)}{5x^3}+ 6/5.\\\qquad \vspace*{-4mm}\\\qquad
\end{equation}
Interestingly, one can see that the above expression \eqref{comocont} is similar to eq. \eqref{TTkernelmd}, and the first term of this expression is identical to the first term of eq. \eqref{TTkernelmdctcom112}. The only oscillating terms $\sin{x}$ or $\cos{x}$ contribute to SIGWs and show the physical behavior of the energy density of SIGWs, and the rest are fictitious.
The evolution of the kernel function \eqref{comocont} in the late time $(x\gg1)$ is  presented in Figures \ref{fig:divergent.MD} and \ref{fig:convergent.MD}. In addition, the constant term in eq. \eqref{TTkernelmdctcom112} does not account for SIGWs, so it makes no contribution to SIGWs if we consider physical SIGWs. On the other hand, we also showed that constant tensor
perturbations in the Poisson gauge do not contribute to the energy density of GWs
even though they appear in the integration kernel. This result in the present work supports our proposal. Moreover,  the evolution for the gauge independent kernel function \eqref{TTkernelmdctcom112} is shown in  Figures \ref{fig:divergent.MDbarringfactor} and \ref{fig:convergent.MDafterbarringfactor},\linebreak respectively.
\subsection{Uniform curvature gauge}\vspace*{-2mm}
The uniform curvature gauge is defined as $\psi=E=0$. One can use the results of the background equations 
in eq. \eqref{I_int}, and after some algebraic manipulations, the kernel in this gauge is given by
\end{multicols}
\noindent\rule{85.5mm}{0.4pt}\rule{0.4pt}{2mm}
\begin{align}
&I_{\text{RD, UC}}(u,v,x)=\,
-\frac{3}{u^3v^3x^4}\left(
(ux)^3v\sin x+u(vx)^3\sin x-3uvx^3\sin x\right)
+\frac{3(u^2+v^2-3)^2}{4u^3v^3x}\times\ln\left[
\Big\lvert \frac{3-(u+v)^2}{3-(u-v)^2}\Big\rvert
\right] \sin x \notag\\&+\frac{3}{4u^3v^3x^4}\left[
-24\left(-ux\cos\frac{ux}{\sqrt{3}}+\sqrt{3}\sin\frac{ux}{\sqrt{3}}\right)\left(-vx\cos\frac{vx}{\sqrt{3}}+\sqrt{3}\sin\frac{vx}{\sqrt{3}}\right)
\right.
-4\left(
+6ux\cos\frac{ux}{\sqrt{3}}
\left(-vx \cos\frac{vx}{\sqrt{3}}+\sqrt{3}\sin\frac{vx}{\sqrt{3}}\right)\right.\nonumber\\
&\left.-3\sin\frac{ux}{\sqrt{3}}\left(
-2\sqrt{3} vx\cos\frac{vx}{\sqrt{3}}+(6+(u^2+v^2-3)x^2)\sin\frac{vx}{\sqrt{3}}
\right)
\right)
-6uvx^2\cos\frac{ux}{\sqrt{3}}\cos\frac{vx}{\sqrt{3}}+6\sqrt{3}ux\cos\frac{ux}{\sqrt{3}}\sin\frac{vx}{\sqrt{3}}\nonumber\\
&+6\sqrt{3}vx\sin\frac{ux}{\sqrt{3}}\cos\frac{vx}{\sqrt{3}}
\left.-18\sin\frac{ux}{\sqrt{3}}\sin\frac{vx}{\sqrt{3}}+(u^2+v^2-3)x^2\sin\frac{ux}{\sqrt{3}}\sin\frac{vx}{\sqrt{3}}
\right]
+(u^2+v^2-3)^2x^3\nonumber\\
&\times\left(
\sin x
\left(
\mathrm{Ci}\left[\left(1+\frac{u-v}{\sqrt{3}}\right)x\right]+
\mathrm{Ci}\left[\left(1+\frac{v-u}{\sqrt{3}}\right)x\right]\right.\right.
\left.\left.-\mathrm{Ci}\left[\left(1+\frac{u+v}{\sqrt{3}}\right)x\right]
-\mathrm{Ci}\left[\left\lvert 1-\frac{u+v}{\sqrt{3}}\right\rvert x\right]
+\ln\left[
\left\lvert \frac{3-(u+v)^2}{3-(u-v)^2}\right\rvert
\right]
\right)\right.\nonumber\\
&
+\cos x\left(
-\mathrm{Si}\left[\left(1+\frac{u-v}{\sqrt{3}}\right)x\right]-
\mathrm{Si}\left[\left(1+\frac{v-u}{\sqrt{3}}\right)x\right]+
\mathrm{Si}\left[\left(1-\frac{u+v}{\sqrt{3}}\right)x\right]\right.
+\left.\left.\left.\mathrm{Si}\left[\left(1+\frac{u+v}{\sqrt{3}}\right)x\right]\right)\right)\right]
+\frac{3(u^2+v^2-3)^2}{4u^3v^3x}\nonumber\\
&\times\left[
\left(
\mathrm{Ci}\left[\left(1+\frac{u-v}{\sqrt{3}}\right)x\right]+
\mathrm{Ci}\left[\left(1+\frac{v-u}{\sqrt{3}}\right)x\right]-
\mathrm{Ci}\left[\left(1+\frac{u+v}{\sqrt{3}}\right)x\right]\right.\right.
\left.
-\mathrm{Ci}\left[\left\lvert 1-\frac{u+v}{\sqrt{3}}\right\rvert x\right]\right)\sin x+\left(
-\mathrm{Si}\left[\left(1+\frac{u-v}{\sqrt{3}}\right)x\right]\right.\nonumber\\
&-\left.\left.
\mathrm{Si}\left[\left(1+\frac{v-u}{\sqrt{3}}\right)x\right]+
\mathrm{Si}\left[\left(1-\frac{u+v}{\sqrt{3}}\right)x\right]+
\mathrm{Si}\left[\left(1+\frac{u+v}{\sqrt{3}}\right)x\right]
\right)\cos x
\right].\label{ucdiver}
\end{align}
\noindent\hspace{92.5mm}\vrule width0.4pt height0.4pt depth 2mm\rule{85.5mm}{0.4pt}
\begin{multicols}{2}
 The above expression \eqref{ucdiver} has some fictitious terms that cause a divergence. To fix this discrepancy,  we show that being second order in perturbation, such induced tensor perturbations are generically gauge independent in contrast to refs. \cite{Hwang:2017oxa, Lu:2020diy, Tomikawa:2019tvi, Ali:2020sfw}, as shown in Figure  \ref{fig:convergentRD}.

Therefore, by using the transformation of a counter term, the kernel function of the gauge independent SIGWs in eq. \eqref{INc} is obtained  to be
\end{multicols}
\noindent\rule{85.5mm}{0.4pt}\rule{0.4pt}{2mm}
\begin{align}
&I_{\text{RD},\,\tilde{h},\, \text{UC}}(u,v,x)
=-\frac{3}{u^3v^3x^4}\left(
(ux)^3v\sin x+u(vx)^3\sin x-3uvx^3\sin x\right)
-\frac{3}{u^3v^3x^4}\Bigg(6uvx^2\cos\frac{ux}{\sqrt{3}}\cos\frac{vx}{\sqrt{3}}+6\sqrt{3}ux\cos\frac{ux}{\sqrt{3}}\sin\frac{vx}{\sqrt{3}}\nonumber\\
&\left.-18\sin\frac{ux}{\sqrt{3}}\sin\frac{vx}{\sqrt{3}}+(u^2+v^2-3)x^2\sin\frac{ux}{\sqrt{3}}\sin\frac{vx}{\sqrt{3}}
\right)
\times\left[
\left(
\mathrm{Ci}\left[\left(1+\frac{u-v}{\sqrt{3}}\right)x\right]+
\mathrm{Ci}\left[\left(1+\frac{v-u}{\sqrt{3}}\right)x\right]-
\mathrm{Ci}\left[\left(1+\frac{u+v}{\sqrt{3}}\right)x\right]\right.\right.\notag\\[5pt]
&
-\mathrm{Ci}\left[\left|1-\frac{u+v}{\sqrt{3}}\right|x\right]\left.
+\ln\left[
\left|\frac{3-(u+v)^2}{3-(u-v)^2}\right|
\right]\right)\sin x+\left(
-\mathrm{Si}\left[\left(1+\frac{u-v}{\sqrt{3}}\right)x\right]-
\mathrm{Si}\left[\left(1+\frac{v-u}{\sqrt{3}}\right)x\right]+
\mathrm{Si}\left[\left(1-\frac{u+v}{\sqrt{3}}\right)x\right]\right.\notag\\[5pt]
&+\left.\left.
\mathrm{Si}\left[\left(1+\frac{u+v}{\sqrt{3}}\right)x\right]
\right)\cos x
\right]
\label{I_uccoun}.
\end{align}
\noindent\hspace{92.5mm}\vrule width0.4pt height0.4pt depth 2mm\rule{85.5mm}{0.4pt}
\begin{multicols}{2}
 The counter-term in the uniform curvature gauge is non-trivial.  Also, as we can see, some terms in the  expression of eq. \eqref{ucdiver} are the same as in eq. \eqref{I_uccoun}. Suppose we had the terms that are freely propagating tensor perturbations and only represented the free GWs. In that case, these free oscillating terms contribute to the energy density $\Omega_{\text{GW}}$ of SIGWs \cite{Inomata:2019yww}.  The evolution of the divergent kernel function $I^2_\text{RD,\,UC}(u,v,x)$ with $u=v=1$ and $u=v=0.1$ is shown in Figure \ref{fig:divergentRD} and for the
gauge independent kernel $I^2_{\tilde{h},\,\text{RD,\,UC}}(u,v,x)$ of the energy density of SIGWs is shown in Figure \ref{fig:convergentRD}, respectively.

In MD, we substitute the results of the transfer functions into the kernel function \eqref{ichi_newtoncase}, and we get
\begin{equation}
\label{Iucmd}
    I_{\text{MD, UC}}(u,v,x)= \frac{18(x\cos x -\sin x)}{5x^3}+\frac{-45x^5+6x^3}{1000x^3}.
\end{equation}
In the MD era, by using eq. \eqref{INc}, we can evaluate the independent  kernel function in the uniform curvature gauge as:
\begin{equation}
\label{TTkcontuc}
I_{\text{MD},\,\tilde{h},\, \text{UC}}(u,v,x)
=\frac{18(x\cos x -\sin x)}{5x^3}+ 6/5.
\end{equation}
One can see that the above expression \eqref{ucdiver} has some extra terms that do not contribute to SIGWs and cause a divergence. However, we find the kernel  with the inclusion of a counter term to remove the divergent term and get the gauge-independent kernel function \eqref{I_uccoun}. Here, the only oscillating terms $\sin{x}$ or $\cos{x}$ contribute to SIGWs and show the physical behavior of the energy density of SIGWs, and the rest are fictitious. Besides, one can see the divergence behavior of the kernel at the subhorizon limit in Figures \ref{fig:divergentRD} and \ref{fig:convergentRD} with $u=v=1$ and $u=v=0.1$, respectively.  While the evolution of the kernel function \eqref{Iucmd} in the late time $(x\gg1)$  is shown in Figures \ref{fig:divergent.MD} and \ref{fig:convergent.MD}, respectively. In addition, the constant term 6/5 in eq. \eqref{TTkcontuc} does not account for SIGWs, so it does not contribute SIGWs if we consider physical SIGWs. On the other hand, we also showed that constant tensor
perturbations in the Poisson gauge do not contribute to the energy density of GWs
even though they appear in the integration kernel. This result in the present work supports our proposal. Moreover,  the evolution for the gauge independent kernel function  \eqref{TTkcontuc}, at the late time limit $(x\gg1)$ is shown in  Figures \ref{fig:divergent.MDbarringfactor} and \ref{fig:convergent.MDafterbarringfactor}, respectively.

\subsection{Total matter gauge}
Next, the total matter gauge is defined by $\delta V=E=0$. For the kernel function \eqref{I_int}, we use some background calculations. After some calculations, one can compute the kernel function $I_{\mathrm{RD,\, TM}}(u,v,x)$ as:
\end{multicols}
\noindent\rule{85.5mm}{0.4pt}\rule{0.4pt}{2mm}
\begin{align}
&I_{\mathrm{RD\,, TM}}(u,v,x)= -\frac{3}{u^3v^3x^4}\left(
(ux)^3v\sin x+u(vx)^3\sin x-3uvx^3\sin x\right)
+\frac{3(u^2+v^2-3)^2}{4u^3v^3x}\times\ln\left[
\Big\lvert \frac{3-(u+v)^2}{3-(u-v)^2}\Big\rvert
\right] \sin x
+\frac{1}{4u^3v^3x^4}\notag\\
&\times \left(
-2\left(
6ux\cos\frac{ux}{\sqrt{3}}+\sqrt{3}(u^2x^2-6)\sin\frac{ux}{\sqrt{3}}
\right)\right.
\times\left(
6vx\cos\frac{vx}{\sqrt{3}}+\sqrt{3}(v^2x^2-6)\sin\frac{vx}{\sqrt{3}}
\right)
-12\left[
6ux\cos\frac{ux}{\sqrt{3}}\left(
-vx\cos\frac{vx}{\sqrt{3}}\right.\right.\nonumber\\&
\left.\left.+\sqrt{3}\sin\frac{vx}{\sqrt{3}}
\right)-3\sin\frac{ux}{\sqrt{3}}\left(
-2\sqrt{3}vx\cos\frac{qx}{\sqrt{3}}+[6+(u^2+v^2-3)^2 x^2]\sin\frac{vx}{\sqrt{3}}\right)\right]
+\frac{3}{4u^3v^3x}\left(
	-\frac{4}{x^3}\left(
	-6uvx^2\cos\frac{ux}{\sqrt{3}}\cos\frac{vx}{\sqrt{3}}\right.\right.\notag\\
	&+6\sqrt{3}ux\cos\frac{ux}{\sqrt{3}}\sin\frac{vx}{\sqrt{3}}+6\sqrt{3}vx\sin\frac{ux}{\sqrt{3}}\cos\frac{vx}{\sqrt{3}}
\left.
-3(6+(u^2+v^2-3)x^2)\sin\frac{ux}{\sqrt{3}}\sin\frac{vx}{\sqrt{3}}
	\right)
+3(u^2+v^2-3)x^3\left[
\sin x\left(\mathrm{Ci}\right.\right.\nonumber\\
&\times\left.\left[\left(1+\frac{u-v}{\sqrt{3}}\right)x\right]+
\mathrm{Ci}\left[\left(1+\frac{v-u}{\sqrt{3}}\right)x\right]-\mathrm{Ci}\left[\left(1+\frac{u+v}{\sqrt{3}}\right)x\right]
-\mathrm{Ci}\left[\left\lvert 1-\frac{u+v}{\sqrt{3}}\right\rvert x\right]
+\ln\left[\left\lvert \frac{3-(u+v)^2}{3-(u-v)^2}\right\rvert \right]\right)
+\cos x\left(-\mathrm{Si}\left[\left(1+\frac{u-v}{\sqrt{3}}\right)x\right]\right.
\notag\\&\left.-
\mathrm{Si}\left[\left(1+\frac{v-u}{\sqrt{3}}\right)x\right]\right.
\left.\left.\left.+\mathrm{Si}\left[\left(1-\frac{u+v}{\sqrt{3}}\right)x\right]+
\mathrm{Si}\left[\left(1+\frac{u+v}{\sqrt{3}}\right)x\right]\right)\right]\right)+(u^2+v^2-3)^2
\times\left[
	\left(
	\mathrm{Ci}\left[\left(1+\frac{u-v}{\sqrt{3}}\right)x\right]+
	\mathrm{Ci}\left[\left(1+\frac{v-u}{\sqrt{3}}\right)x\right]\right.\right.\notag\\
	&\left.-
	\mathrm{Ci}\left[\left(1+\frac{u+v}{\sqrt{3}}\right)x\right]
	-\mathrm{Ci}\left[\left\lvert 1-\frac{u+v}{\sqrt{3}}\right\rvert x\right]
	\right)\sin x+\left(
	-\mathrm{Si}\left[\left(1+\frac{u-v}{\sqrt{3}}\right)x\right]-
	\mathrm{Si}\left[\left(1+\frac{v-u}{\sqrt{3}}\right)x\right]\right.\notag\\&
\left.\left.\left.+
	\mathrm{Si}\left[\left(1-\frac{u+v}{\sqrt{3}}\right)x\right]+
	\mathrm{Si}\left[\left(1+\frac{u+v}{\sqrt{3}}\right)x\right]
	\right)\cos x
	\right]
	\right)\label{tmieq}.
\end{align}
\noindent\hspace{92.5mm}\vrule width0.4pt height0.4pt depth 2mm\rule{85.5mm}{0.4pt}
\begin{multicols}{2}
 It is to be noted that the above expression \eqref{tmieq} is different from the kernel in the Poisson gauge. There exist some extra terms in eq. \eqref{tmieq}.  Moreover, one can see the behavior of the evolution of \eqref{tmieq} at the subhorizon limit in Figure \ref{fig:divergentRD} with $u=v=1$ and $u=v=0.1$. Here, it is vital to present a resolution to these discrepancies by using a counter term given in the aforementioned expression \eqref{hchi}. We show that being second order in perturbation, such induced tensor perturbations are generically gauge independent in contrast to refs. \cite{Hwang:2017oxa, Lu:2020diy, Tomikawa:2019tvi, Ali:2020sfw}, as shown in Figure  \ref{fig:convergentRD} with $u=v=1$ and $u=v=0.1$.

Therefore, by using the transformation of a counter term, the kernel function of the gauge independent SIGWs in eq. \eqref{INc} is obtained  to be
\end{multicols}
\noindent\rule{85.5mm}{0.4pt}\rule{0.4pt}{2mm}
\begin{align}
&I_{\text{RD},\,\tilde{h},\, \text{TM}}(u,v,x)
=-\frac{3}{u^3v^3x^4}\left(
(ux)^3v\sin x+u(vx)^3\sin x-3uvx^3\sin x\right)
-\frac{3}{u^3v^3x^4}\Bigg(6uvx^2\cos\frac{ux}{\sqrt{3}}\cos\frac{vx}{\sqrt{3}}+6\sqrt{3}ux\cos\frac{ux}{\sqrt{3}}\sin\frac{vx}{\sqrt{3}}\nonumber\\
&\left.-18\sin\frac{ux}{\sqrt{3}}\sin\frac{vx}{\sqrt{3}}+(u^2+v^2-3)x^2\sin\frac{ux}{\sqrt{3}}\sin\frac{vx}{\sqrt{3}}
\right)
\times\left[
\left(
\mathrm{Ci}\left[\left(1+\frac{u-v}{\sqrt{3}}\right)x\right]+
\mathrm{Ci}\left[\left(1+\frac{v-u}{\sqrt{3}}\right)x\right]-
\mathrm{Ci}\left[\left(1+\frac{u+v}{\sqrt{3}}\right)x\right]\right.\right.\notag\\
&\left.
-\mathrm{Ci}\left[\left|1-\frac{u+v}{\sqrt{3}}\right|x\right]
+\ln\left[
\left|\frac{3-(u+v)^2}{3-(u-v)^2}\right|
\right]\right)\sin x+\left(
-\mathrm{Si}\left[\left(1+\frac{u-v}{\sqrt{3}}\right)x\right]\right.-
\mathrm{Si}\left[\left(1+\frac{v-u}{\sqrt{3}}\right)x\right]+
\mathrm{Si}\left[\left(1-\frac{u+v}{\sqrt{3}}\right)x\right]\notag\\&
\left.\left.+
\mathrm{Si}\left[\left(1+\frac{u+v}{\sqrt{3}}\right)x\right]
\right)\cos x
\right]
\label{I_Pcomo123}.
\end{align}
\noindent\hspace{92.5mm}\vrule width0.4pt height0.4pt depth 2mm\rule{85.5mm}{0.4pt}
\begin{multicols}{2}
 The counter-term in the total matter gauge is non-trivial.
Besides, one can see that the terms in the expression \eqref{tmieq} are extra compared to eq. \eqref{I_Pcomo123}, which cause discrepancies in the total matter gauge during RD.  Suppose we had the terms that are freely propagating tensor perturbations and only represented the free GWs. In that case, these free oscillating terms contribute to the energy density $\Omega_{\text{GW}}$ of SIGWs \cite{Inomata:2019yww}.  The evolution of the divergent kernel function $I^2_\text{RD,\,TM}(u,v,x)$ with $u=v=1$ and $u=v=0.1$ is shown in Figure \ref{fig:divergentRD} and for the
gauge independent kernel $I^2_{\tilde{h},\,\text{RD,\,TM}}(u,v,x)$ of the energy density of SIGWs is shown in Figure \ref{fig:convergentRD}, respectively.

In MD universe, to evaluate the kernel function,  we use 
 eq. \eqref{ichi_newtoncase}, and find
\begin{align}
I_{\text{MD, \,TM}}(u,v,x)=\, \frac{18(x\cos x -\sin x)}{5x^3}+
\frac{-5x^5+6x^3}{2500x^3}.
\label{MD_tmmd}
\end{align}
Now in the MD era, by using eq. \eqref{ichi_newtoncaseXi}, we can evaluate the gauge-independent  kernel function in the total matter gauge as:
\begin{equation}
I_{\text{MD},\,\tilde{h},\, \text{TM}}(u,v,x)
=\frac{18(x\cos x -\sin x)}{5x^3}+ 6/5.
\label{TTkernelmdctcom}
\end{equation}
Here, one can see that the first term of the expression \eqref{MD_tmmd} is identical to the first term of eq. \eqref{TTkernelmdctcom}. The only oscillating terms $\sin{x}$ or $\cos{x}$ contribute to SIGWs and show the physical behavior of the energy density of SIGWs, and the rest are fictitious.
The evolution of the kernel function \eqref{comocont} in the late time $(x\gg1)$ is  presented in Figures \ref{fig:divergent.MD} and \ref{fig:convergent.MD}. In addition, the constant term in eq. \eqref{TTkernelmdctcom} does not account for SIGWs, so it makes no contribution to SIGWs if we consider physical SIGWs. On the other hand, we also showed that constant tensor
perturbations in the Poisson gauge do not contribute to the energy density of GWs even though they appear in the integration kernel. This result in the present work supports our proposal. Moreover,  the evolution for the gauge independent kernel function \eqref{TTkernelmdctcom} is shown in  Figures \ref{fig:divergent.MDbarringfactor} and \ref{fig:convergent.MDafterbarringfactor}, respectively.
\subsection{Uniform density gauge}
The uniform density gauge is defined by $ \delta\rho=E=0$. To evaluate the kernel function in this gauge, after some straightforward calculations, we have
\end{multicols}
\noindent\rule{85.5mm}{0.4pt}\rule{0.4pt}{2mm}
\begin{align}
&I_\text{RD, UD}(u,v,x)=\,-\frac{3}{u^3v^3x^4}\left(
(ux)^3v\sin x+u(vx)^3\sin x-3uvx^3
\sin x\right)+\frac{3(u^2+v^2-3)^2}{4u^3v^3x}\times\ln\left[
\Big\lvert \frac{3-(u+v)^2}{3-(u-v)^2}\Big\rvert
\right] \sin x 
 +\frac{1}{4 u^3 v^3 x^4}\notag\\
 &\times
 \left[
2 u x \left(u^2 x^2\!-\!6\right) \cos \left(\!\!\frac{u x}{\sqrt{3}}\!\right)\left(v x \left(v^2 x^2\!-\!6\right) \cos
\left(\frac{v x}{\sqrt{3}}\right)\right.\right.
\left.
\!-\!2 \sqrt{3} \left(v^2 x^2\!-\!3\right) \sin \left(\!\!\frac{v x}{\sqrt{3}}\!\right)\right)
\!-\!4 \left(u^2 x^2\!-\!3\right) \sin \left(\frac{u x}{\sqrt{3}}\right)
\times\!\!\left.\left(\!\!\sqrt{3} v x \left(v^2 x^2\!-\!6\right) \cos \left(\!\!\frac{v x}{\sqrt{3}}\right)\right.\right.\notag\\
&\left.\left.-6 \left(v^2 x^2-3\right) \sin \left(\frac{v x}{\sqrt{3}}\right)\right)
\right]
+\frac{3}{4u^3v^3x}\left(
	-\frac{4}{x^3}\left(
	-6uvx^2\cos\frac{ux}{\sqrt{3}}\cos\frac{vx}{\sqrt{3}}\right.\right.
+6\sqrt{3}ux\cos\frac{ux}{\sqrt{3}}\sin\frac{vx}{\sqrt{3}}+6\sqrt{3}vx\sin\frac{ux}{\sqrt{3}}\cos\frac{vx}{\sqrt{3}}\notag\\
	&\left.-3(6+(u^2+v^2-3)x^2)\sin\frac{ux}{\sqrt{3}}\sin\frac{vx}{\sqrt{3}}
	\right)+(u^2+v^2-3)^2
\times\left[
	\left(
	\mathrm{Ci}\left[\left(1+\frac{u-v}{\sqrt{3}}\right)x\right]+
	\mathrm{Ci}\left[\left(1+\frac{v-u}{\sqrt{3}}\right)x\right]-
	\mathrm{Ci}\left[\left(1+\frac{u+v}{\sqrt{3}}\right)x\right]\right.\right.\notag\\
	&\left.
	-\mathrm{Ci}\left[\lvert 1-\frac{u+v}{\sqrt{3}}\rvert x\right]
	\right)\sin x+\left(
	-\mathrm{Si}\left[\left(1+\frac{u-v}{\sqrt{3}}\right)x\right]-
	\mathrm{Si}\left[\left(1+\frac{v-u}{\sqrt{3}}\right)x\right]\right.
\left.\left.\left.+
	\mathrm{Si}\left[\left(1-\frac{u+v}{\sqrt{3}}\right)x\right]+
	\mathrm{Si}\left[\left(1+\frac{u+v}{\sqrt{3}}\right)x\right]
	\right)\cos x
	\right]
	\right)\label{IalphaUD}.
\end{align}
\noindent\hspace{92.5mm}\vrule width0.4pt height0.4pt depth 2mm\rule{85.5mm}{0.4pt}
\begin{multicols}{2}
In the above expression \eqref{IalphaUD}, we see that there are extra terms when we compare it with the result of the kernel in the Poisson gauge.  This different result with extra terms causes divergence. This work shows that they lead to discrepancies between the Poisson gauge and the uniform density gauge.  Here, we try to present a resolution by making use of a counter term given in the aforementioned expression \eqref{hchi}. We show that being second order in perturbation, such induced tensor perturbations are generically gauge independent in contrast to refs. \cite{Hwang:2017oxa, Lu:2020diy, Tomikawa:2019tvi, Ali:2020sfw}, as shown in Figure  \ref{fig:convergentRD}.

Therefore, by using the transformation of a counter term, the kernel function of the gauge independent SIGWs in eq. \eqref{INc} is obtained  to be
\end{multicols}
\noindent\rule{85.5mm}{0.4pt}\rule{0.4pt}{2mm}
\begin{align}
&I_{\text{RD},\,\tilde{h},\, \text{UD}}(u,v,x)
=-\frac{3}{u^3v^3x^4}\left(
(ux)^3v\sin x+u(vx)^3\sin x-3uvx^3\sin x\right)
-\frac{3}{u^3v^3x^4}\Bigg(6uvx^2\cos\frac{ux}{\sqrt{3}}\cos\frac{vx}{\sqrt{3}}+6\sqrt{3}ux\cos\frac{ux}{\sqrt{3}}\sin\frac{vx}{\sqrt{3}}\nonumber\\
&\left.-18\sin\frac{ux}{\sqrt{3}}\sin\frac{vx}{\sqrt{3}}+(u^2+v^2-3)x^2\sin\frac{ux}{\sqrt{3}}\sin\frac{vx}{\sqrt{3}}
\right)
\times\left[
\left(
\mathrm{Ci}\left[\left(1+\frac{u-v}{\sqrt{3}}\right)x\right]+
\mathrm{Ci}\left[\left(1+\frac{v-u}{\sqrt{3}}\right)x\right]-
\mathrm{Ci}\left[\left(1+\frac{u+v}{\sqrt{3}}\right)x\right]\right.\right.\notag\\
&\left.
-\mathrm{Ci}\left[\left|1-\frac{u+v}{\sqrt{3}}\right|x\right]
+\ln\left[
\left|\frac{3-(u+v)^2}{3-(u-v)^2}\right|
\right]\right)\sin x+\left(
-\mathrm{Si}\left[\left(1+\frac{u-v}{\sqrt{3}}\right)x\right]\right.-
\mathrm{Si}\left[\left(1+\frac{v-u}{\sqrt{3}}\right)x\right]+
\mathrm{Si}\left[\left(1-\frac{u+v}{\sqrt{3}}\right)x\right]\notag\\&
\left.\left.+
\mathrm{Si}\left[\left(1+\frac{u+v}{\sqrt{3}}\right)x\right]
\right)\cos x
\right]
\label{iudens}.
\end{align}
\noindent\hspace{92.5mm}\vrule width0.4pt height0.4pt depth 2mm\rule{85.5mm}{0.4pt}
\begin{multicols}{2}
 The counter-term in the uniform density gauge is non-trivial.
Besides, it can be seen from the evolution of the expression of the kernel \eqref{IalphaUD}  that it is different from the Poisson gauge and has divergence that will break down the perturbation theory \cite{Lu:2020diy, Ali:2020sfw}. However, we try to fix this discrepancy and use the transformation \eqref{gaugetranf1} having counter term $\Xi_{kl}$. We find the gauge-independent kernel function \eqref{iudens} in the uniform density gauge. In addition, it is supposed we had the terms freely propagating tensor perturbations and only represented the free GWs. In that case, these free oscillating terms contribute to the energy density $\Omega_{\text{GW}}$ of SIGWs \cite{Inomata:2019yww}.  The evolution of the divergent kernel function $I^2_\text{RD,\,UD}(u,v,x)$ with $u=v=1$ and $u=v=0.1$ is shown in Figure \ref{fig:divergentRD} and for the
gauge independent kernel $I^2_{\tilde{h},\,\text{RD,\,UD}}(u,v,x)$ of the energy density of SIGWs is shown in Figure \ref{fig:convergentRD}, respectively.

In the MD universe, from eq. \eqref{ichi_newtoncase}, we find the kernel function as follows:
\begin{align}
\label{IalphaUDM}
  I_{\text{MD, \,UD}}&(u,v,x)=\frac{18(x\cos x -\sin x)}{5x^3}\notag\\
  &\qquad +\frac{-x^5 (12+u^2 x^2) (12+v^2x^2)+216000x^3}{36000x^3}.\!\!\!
  \end{align}
While during MD, by using eq. \eqref{INc}, we can evaluate the gauge independent  kernel function in the uniform density gauge as:
\begin{equation}
\label{TTkernelunden}
I_{\text{MD},\,\tilde{h},\, \text{UD}}(u,v,x)
=\frac{18(x\cos x -\sin x)}{5x^3}+ 6/5.
\end{equation}
Here, we can see that the first term in the above expression \eqref{IalphaUDM} is similar to the oscillating term of eq. \eqref{TTkernelunden}. The only oscillating terms $\sin{x}$ or $\cos{x}$ contribute to SIGWs and show the physical behavior of the energy density of SIGWs, and the rest are fictitious.
The evolution of the kernel function \eqref{IalphaUDM} in the late time $(x\gg1)$ is presented in Figures \ref{fig:divergent.MD} and \ref{fig:convergent.MD}. In addition, the constant term in eq. \eqref{TTkernelunden} does not account for SIGWs, so it does not contribute SIGWs if we consider physical SIGWs. On the other hand, we also showed that constant tensor
perturbations in the Poisson gauge do not contribute to the energy density of GWs
even though they appear in the integration kernel. This result in the present work supports our proposal. Moreover,  the evolution for the gauge independent kernel function \eqref{TTkernelunden} is shown in  Figures \ref{fig:divergent.MDbarringfactor} and \ref{fig:convergent.MDafterbarringfactor}, respectively.
\vspace*{-3mm}
\subsection{Uniform expansion gauge}
\begin{figure*}[!b]
\centering
\includegraphics[width=.46\linewidth]{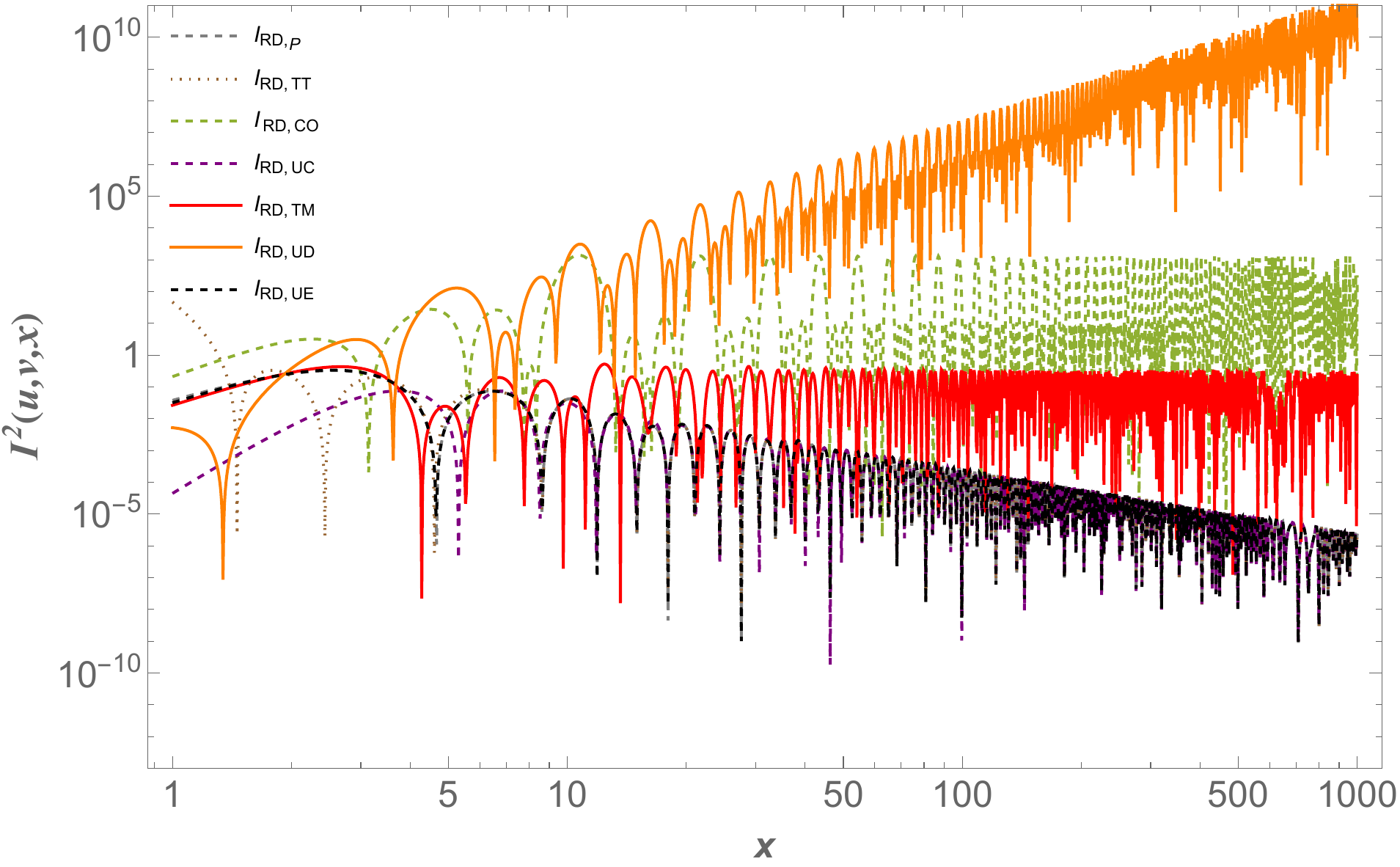} \qquad
\includegraphics[width=.46\linewidth]{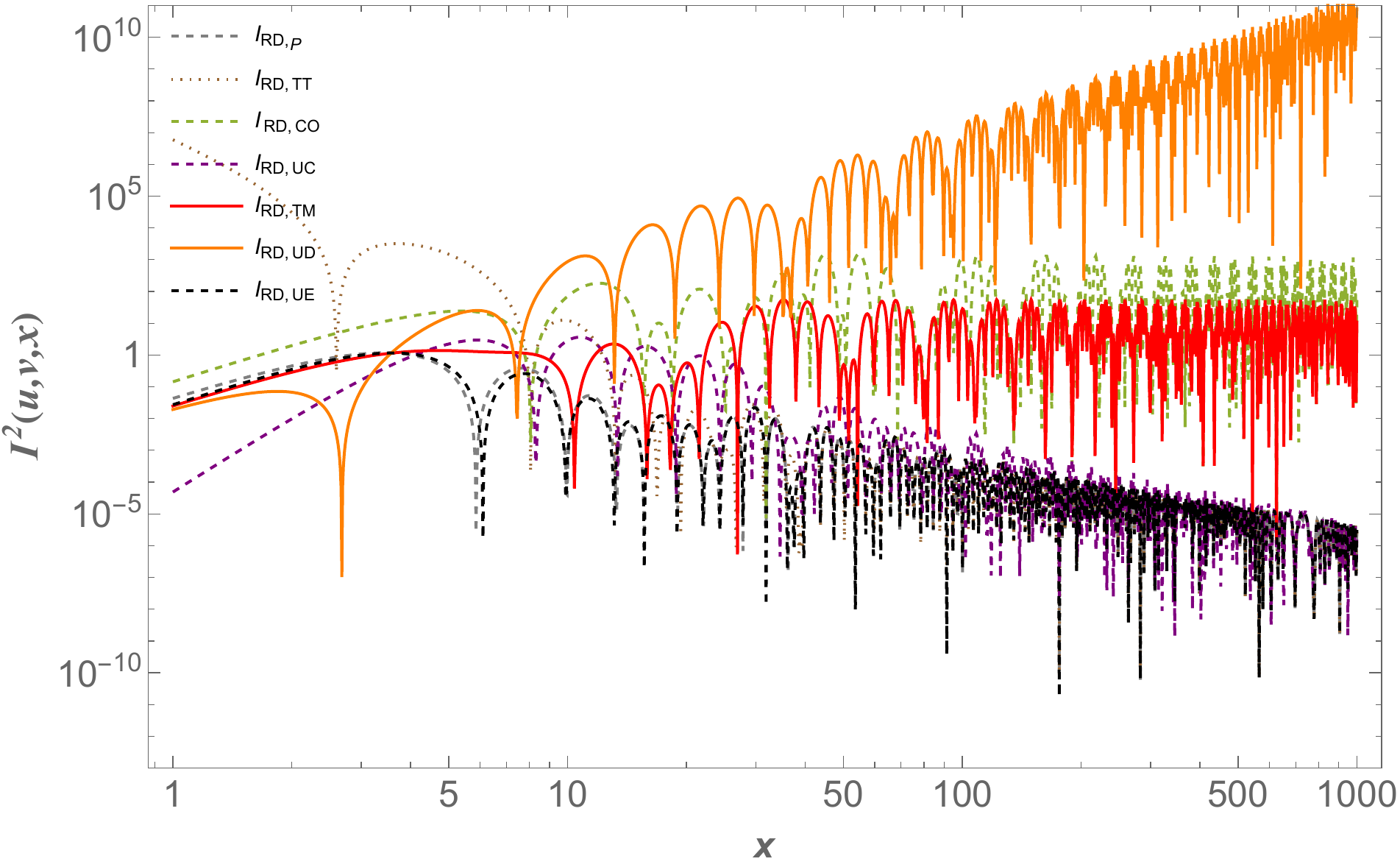}\vspace*{-1mm}
    \caption{(Color online) The evolution of $\lvert I_{\text{RD},P}\rvert$, $\lvert I_{\text{RD},\text{TT}}\rvert$, $\lvert I_{\text{RD},\text{CO}}\rvert$, $\lvert I_{\text{RD},\text{UC}}\rvert$, $\lvert I_{\text{RD},\text{TM}}\rvert$, $\lvert I_{\text{RD},\text{UD}}\rvert$,  and $\lvert I_{\text{RD,UE}}\rvert$ is shown to be divergent at subhorizon limit in some gauges. {Left:} we take $u=v=1.$ {Right:} we let $u=1$ {and}  $v=0.1$.}\label{fig:divergentRD}
\end{figure*}

\begin{figure*}[!b]
\centering\vspace*{3mm}
\includegraphics[width=.46\linewidth]{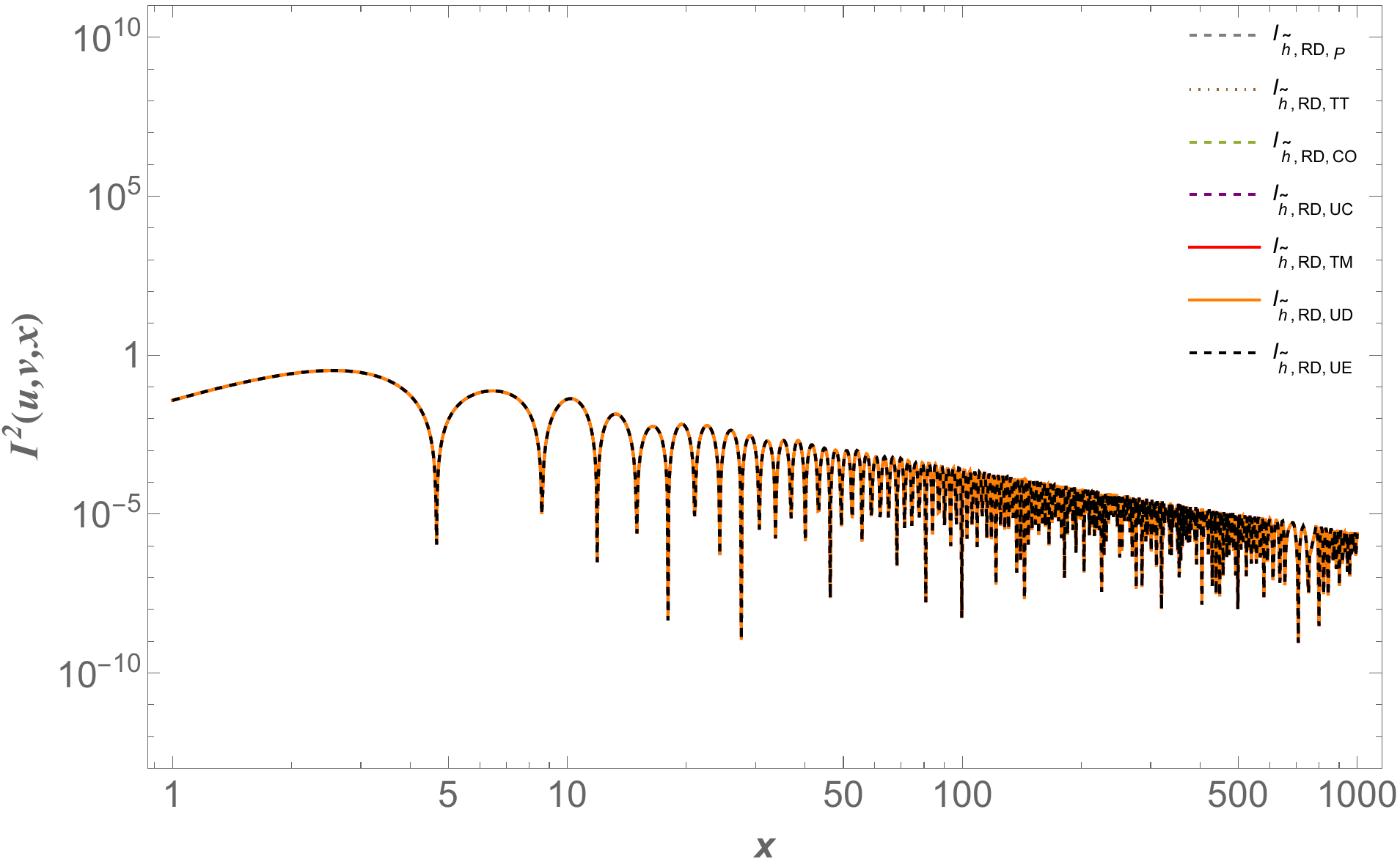} \qquad
\includegraphics[width=.46\linewidth]{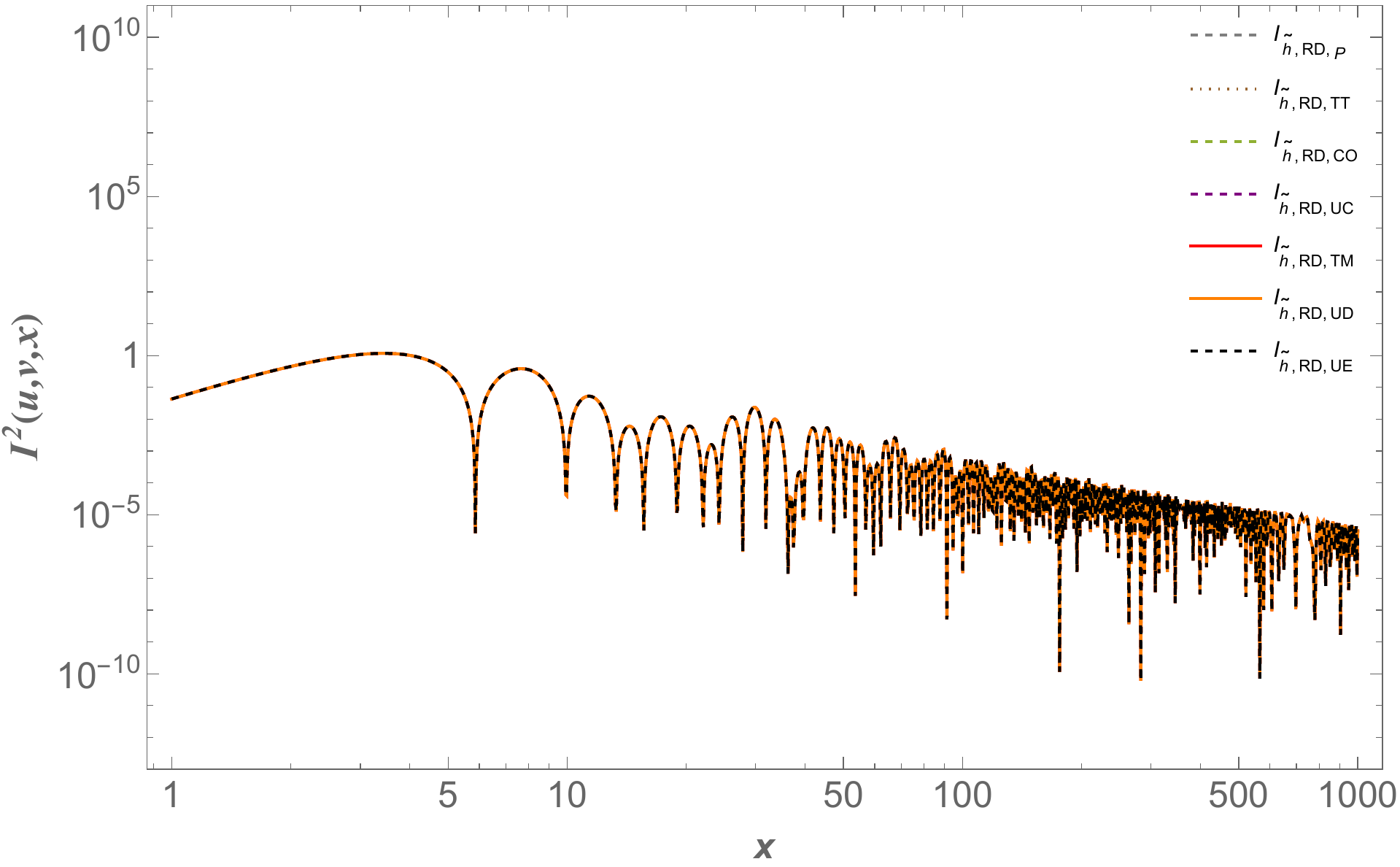}%
    \caption{(Color online) The evolution of $\lvert I_{\tilde{h},\text{RD},\, P}\rvert$, $\lvert I_{\tilde{h},\,\text{RD},\text{TT}}\rvert$, $\lvert I_{\tilde{h},\,\text{RD},\text{CO}}\rvert $, $\lvert I_{\tilde{h},\,\text{RD},\text{UC}}\rvert $, $\lvert I_{\tilde{h},\,\text{RD},\text{TM}}\rvert$, $\lvert I_{\tilde{h},\,\text{RD},\text{UD}}\rvert $,  and $\lvert I_{\tilde{h},\,\text{RD, UE}}\rvert$ is  shown to be convergent at subhorizon limit in all of the gauges. {Left:} we take $u=v=1.$ {Right:} we let $u=1$ {and}  $v=0.1$.}
    \label{fig:convergentRD}%
\end{figure*}

\vspace*{-2mm}
Finally, we consider the uniform expansion gauge, defined by $3(\mathcal{H}\phi+\psi^\prime)+k^2\sigma=0, E=0 $. From  eq. \eqref{ichi_newton}, we can calculate the kernel function in this gauge.
After some algebraic manipulations, we get
\end{multicols}
\noindent\rule{85.5mm}{0.4pt}\rule{0.4pt}{2mm}
\begin{align}
&I_{\text{RD},\, \text{UE}}(u,v,x)=-\frac{3}{u^3v^3x^4}\left(
(ux)^3v\sin x+u(vx)^3\sin x-3uvx^3\sin x\right)
+\frac{3(u^2+v^2-3)^2}{4u^3v^3x}\times\ln\left[
\Big\lvert \frac{3-(u+v)^2}{3-(u-v)^2}\Big\rvert
\right] \sin x \notag\\&
+\frac{9}{u^3 q^3 x^4 \left(u^2 x^2+6\right) \left(v^2 x^2+6\right)}\left[
u^2 v^2 x^4 \sin \left(\frac{u x}{\sqrt{3}}\right) \sin \left(\frac{v x}{\sqrt{3}}\right)\right.
+12 \sqrt{3} u^2 v x^3 \sin \left(\frac{u x}{\sqrt{3}}\right) \cos \left(\frac{v x}{\sqrt{3}}\right)-36 u^2 x^2 \sin \left(\frac{u x}{\sqrt{3}}\right) \sin \left(\frac{v x}{\sqrt{3}}\right)\nonumber\\
&
+12 \sqrt{3} u v^2 x^3 \cos \left(\frac{u x}{\sqrt{3}}\right) \sin \left(\frac{v x}{\sqrt{3}}\right)-6 v^2 x^2 \sin \left(\frac{u x}{\sqrt{3}}\right) \sin \left(\frac{v x}{\sqrt{3}}\right)
+72 u v x^2 \cos \left(\frac{u x}{\sqrt{3}}\right) \cos \left(\frac{v x}{\sqrt{3}}\right)+116 \sin \left(\frac{u x}{\sqrt{3}}\right) \sin \left(\frac{v x}{\sqrt{3}}\right)\nonumber\\
&\left.-72 \sqrt{3} v x \sin \left(\frac{u x}{\sqrt{3}}\right) \cos \left(\frac{v x}{\sqrt{3}}\right)
-72 \sqrt{3} u x \cos \left(\frac{u x}{\sqrt{3}}\right) \sin \left(\frac{v x}{\sqrt{3}}\right)\right]
-\frac{3}{u^3v^3x^4}\left(
-6uvx^2\cos\frac{ux}{\sqrt{3}}\cos\frac{vx}{\sqrt{3}}+6\sqrt{3}ux\cos\frac{ux}{\sqrt{3}}\sin\frac{vx}{\sqrt{3}}\right.\notag\\&
+6\sqrt{3}vx\sin\frac{ux}{\sqrt{3}}\cos\frac{vx}{\sqrt{3}}-18\sin\frac{ux}{\sqrt{3}}\sin\frac{vx}{\sqrt{3}}
\left.+(u^2+v^2-3)x^2\sin\frac{ux}{\sqrt{3}}\sin\frac{vx}{\sqrt{3}}
\right)+\frac{3(u^2+v^2-3)^2}{4u^3v^3x}
\times\left[
\left(
\mathrm{Ci}\left[\left(1+\frac{u-v}{\sqrt{3}}\right)x\right]\right.\right.\nonumber\\
&\left.+
\mathrm{Ci}\left[\left(1+\frac{v-u}{\sqrt{3}}\right)x\right]-
\mathrm{Ci}\left[\left(1+\frac{u+v}{\sqrt{3}}\right)x\right]
-\mathrm{Ci}\left[\left\lvert 1-\frac{u+v}{\sqrt{3}}\right\rvert x\right]
\right)\sin x+\left(
-\mathrm{Si}\left[\left(1+\frac{u-v}{\sqrt{3}}\right)x\right]-
\mathrm{Si}\left[\left(1+\frac{v-u}{\sqrt{3}}\right)x\right]\right.\nonumber\\
&\left.\left.+
\mathrm{Si}\left[\left(1-\frac{u+v}{\sqrt{3}}\right)x\right]+
\mathrm{Si}\left[\left(1+\frac{u+v}{\sqrt{3}}\right)x\right]
\right)\cos x
\right]\label{IalphaUniexp}.
\end{align}
\noindent\hspace{92.5mm}\vrule width0.4pt height0.4pt depth 2mm\rule{85.5mm}{0.4pt}
\begin{multicols}{2}
Similar to the discussion in the previous subsection, we find the kernel \eqref{IalphaUniexp} in the uniform expansion gauge.
We show the evolution of kernel \eqref{IalphaUniexp}  in Figure \ref{fig:divergentRD}. Furthermore, we try to present a resolution by making use of a counter term given in the aforementioned expression \eqref{hchi}. We show that being second order in perturbation, such induced tensor perturbations are generically gauge independent in contrast to refs. \cite{Hwang:2017oxa, Lu:2020diy, Tomikawa:2019tvi, Ali:2020sfw}, as shown in Figure  \ref{fig:convergentRD}.
Therefore, by using the transformation of a counter term, the kernel function of the gauge independent SIGWs in eq. \eqref{INc} is obtained  to be
\end{multicols}
\noindent\rule{85.5mm}{0.4pt}\rule{0.4pt}{2mm}
\begin{align}
&I_{\text{RD},\,\tilde{h},\, \text{UE}}(u,v,x)
=-\frac{3}{u^3v^3x^4}\left(
(ux)^3v\sin x+u(vx)^3\sin x-3uvx^3\sin x\right)
-\frac{3}{u^3v^3x^4}\Bigg(6uvx^2\cos\frac{ux}{\sqrt{3}}\cos\frac{vx}{\sqrt{3}}+6\sqrt{3}ux\cos\frac{ux}{\sqrt{3}}\sin\frac{vx}{\sqrt{3}}\nonumber\\
&\left.-18\sin\frac{ux}{\sqrt{3}}\sin\frac{vx}{\sqrt{3}}+(u^2+v^2-3)x^2\sin\frac{ux}{\sqrt{3}}\sin\frac{vx}{\sqrt{3}}
\right)
\times\left[
\left(
\mathrm{Ci}\left[\left(1+\frac{u-v}{\sqrt{3}}\right)x\right]+
\mathrm{Ci}\left[\left(1+\frac{v-u}{\sqrt{3}}\right)x\right]-
\mathrm{Ci}\left[\left(1+\frac{u+v}{\sqrt{3}}\right)x\right]\right.\right.\notag\\
&\left.
-\mathrm{Ci}\left[\left|1-\frac{u+v}{\sqrt{3}}\right|x\right]
+\ln\left[
\left|\frac{3-(u+v)^2}{3-(u-v)^2}\right|
\right]\right)\sin x+\left(
-\mathrm{Si}\left[\left(1+\frac{u-v}{\sqrt{3}}\right)x\right]\right.-
\mathrm{Si}\left[\left(1+\frac{v-u}{\sqrt{3}}\right)x\right]+
\mathrm{Si}\left[\left(1-\frac{u+v}{\sqrt{3}}\right)x\right]\notag\\&
\left.\left.+
\mathrm{Si}\left[\left(1+\frac{u+v}{\sqrt{3}}\right)x\right]
\right)\cos x
\right]
.\label{I_Pcomo12}
\end{align}
\noindent\rule{85.5mm}{0.4pt}\rule{0.4pt}{2mm}
\begin{multicols}{2}
 The counter-term in the uniform expansion gauge is non-trivial.
Besides, as we can see, the terms in the first two lines of the expression \eqref{ICOM} are the same with eq. \eqref{I_Pcomo12}. Suppose we had the terms that are freely propagating tensor perturbations and only represented the free GWs. In that case, these free oscillating terms contribute to the energy density $\Omega_{\text{GW}}$ of SIGWs \cite{Inomata:2019yww}.  The evolution of the divergent kernel function $I^2_\text{RD,\,UE}(u,v,x)$ with $u=v=1$ and $u=v=0.1$ is shown in Figure \ref{fig:divergentRD} and for the
gauge independent kernel $I^2_{\tilde{h},\,\text{RD,\,UE}}(u,v,x)$ of the energy density of SIGWs is shown in Figure \ref{fig:convergentRD}, respectively.

In the MD universe, from eq. \eqref{ichi_newtoncase}, one can
obtain the result of kernel function as: \begin{align}
\label{IalphaUnieUD}
  I_{\text{{MD,\, UE}}}&(u,v,x)=\frac{18(x\cos x -\sin x)}{5x^3}\notag\\&\qquad+\frac{-810 x^5+150x^3(u^2 x^2+18) (v^2 x^2+18)}{125x^3(u^2 x^2+18) (v^2 x^2+18)}.\!
\end{align}
For MD, with the use of eq. \eqref{INc}, one can calculate  the gauge independent  kernel function in the uniform expansion gauge as:
\begin{equation}
\label{TTkernelmduex}
I_{\text{MD},\,\tilde{h},\, \text{UE}}(u,v,x)
=\frac{18(x\cos x -\sin x)}{5x^3}+ 6/5.
\end{equation}
Here, one can see that the first term in the above expression \eqref{IalphaUnieUD} is appearing in  eq. \eqref{TTkernelmduex}. The only oscillating terms $\sin{x}$ or $\cos{x}$ contribute to SIGWs and show the physical behavior of the energy density of SIGWs, and the rest are fictitious.
The evolution of the kernel function \eqref{IalphaUnieUD} in the late time $(x\gg1)$ is presented in Figures \ref{fig:divergent.MD} and \ref{fig:convergent.MD}. In addition, the constant term in eq. \eqref{TTkernelmduex} does not account for SIGWs, so it makes no contribution to SIGWs if we consider physical SIGWs. On the other hand, we also showed that constant tensor
perturbations in the Poisson gauge do not contribute to the energy density of GWs
even though they appear in the integration kernel. This result in the present work supports our proposal. Moreover,  the evolution for the gauge independent kernel function \eqref{TTkernelmduex} is shown in  Figures \ref{fig:divergent.MDbarringfactor} and \ref{fig:convergent.MDafterbarringfactor}, respectively.

To examine the energy density $\Omega_{\text{GW}}$ of SIGWs in various gauge choices, the secondary scalar-induced tensor is of two forms. One kind of tensor perturbation is freely propagating tensor perturbations whose oscillations are like $\sin(k \eta)$ or $\cos(k \eta)$. This kind of perturbation contributes to SIGWs.
The other tensor perturbations with terms instead of $\sin(k \eta)$ or $\cos(k \eta)$ do not contribute to SIGWs. In the following, we will shed light on different gauge choices and confirm with comparison among the kernel functions whether SIGWs are gauge (in)dependent in different gauges with the above identification of SIGWs.
\section{Comparison among the kernel functions in various gauges}
\label{sec.3case}
\begin{figure*}[!b]
\centering
 \includegraphics[width=.46\linewidth]{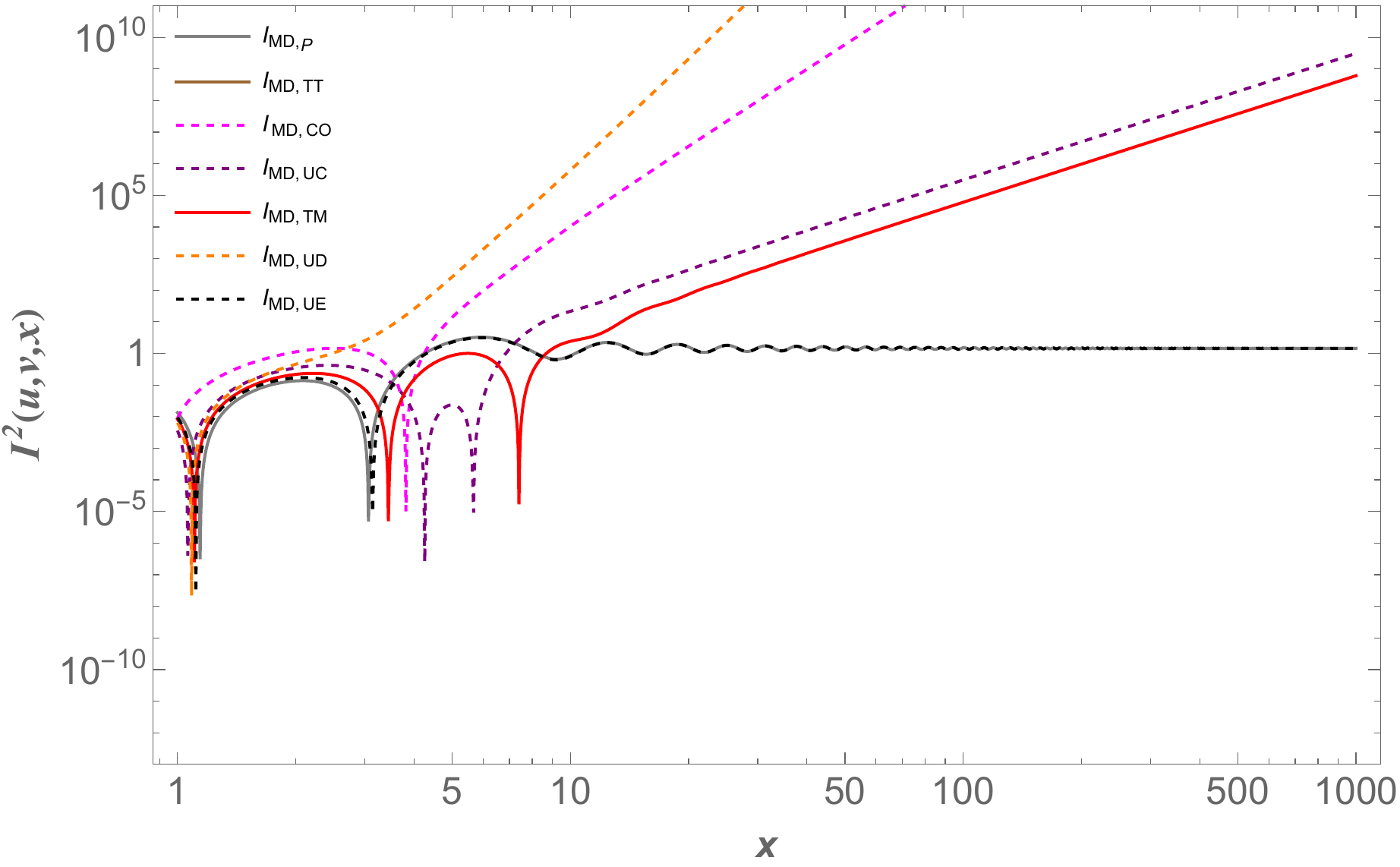} \qquad
    \includegraphics[width=.46\linewidth]{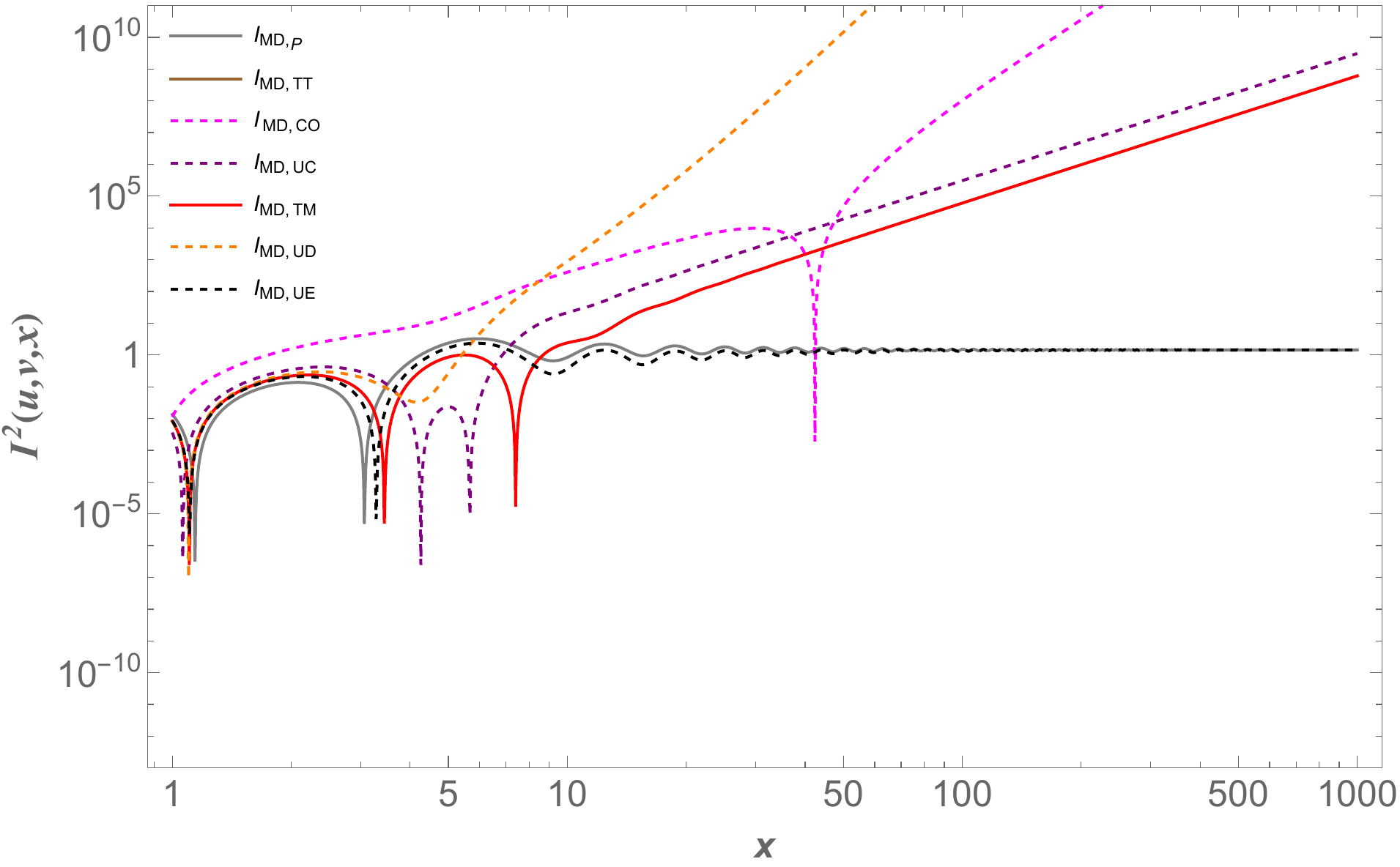}%
    \caption{(Color online) The evolution of $\lvert I_{\text{MD},P}\rvert $, $\lvert I_{\text{MD},\text{TT}}\rvert $, $\lvert I_{\text{}\text{MD},\text{CO}}\rvert $, $\lvert I_{\text{MD},\text{UC}}\rvert$, $\lvert I_{\text{MD},\text{TM}}\rvert$, $\lvert I_{\text{MD},\text{UD}}\rvert$,  and $\lvert I_{\text{MD,UE}}\rvert $ is  shown to be divergent at the subhorizon limit in different gauges. Here, we show kernels with a constant factor 6/5. {{Left:}} we take $u=v=1.$ {{Right:}} we let $u=1$ {and}  $v=0.1$.}\label{fig:divergent.MD}
\end{figure*}

\begin{figure*}[!b]
\centering
  \includegraphics[width=.46\linewidth]{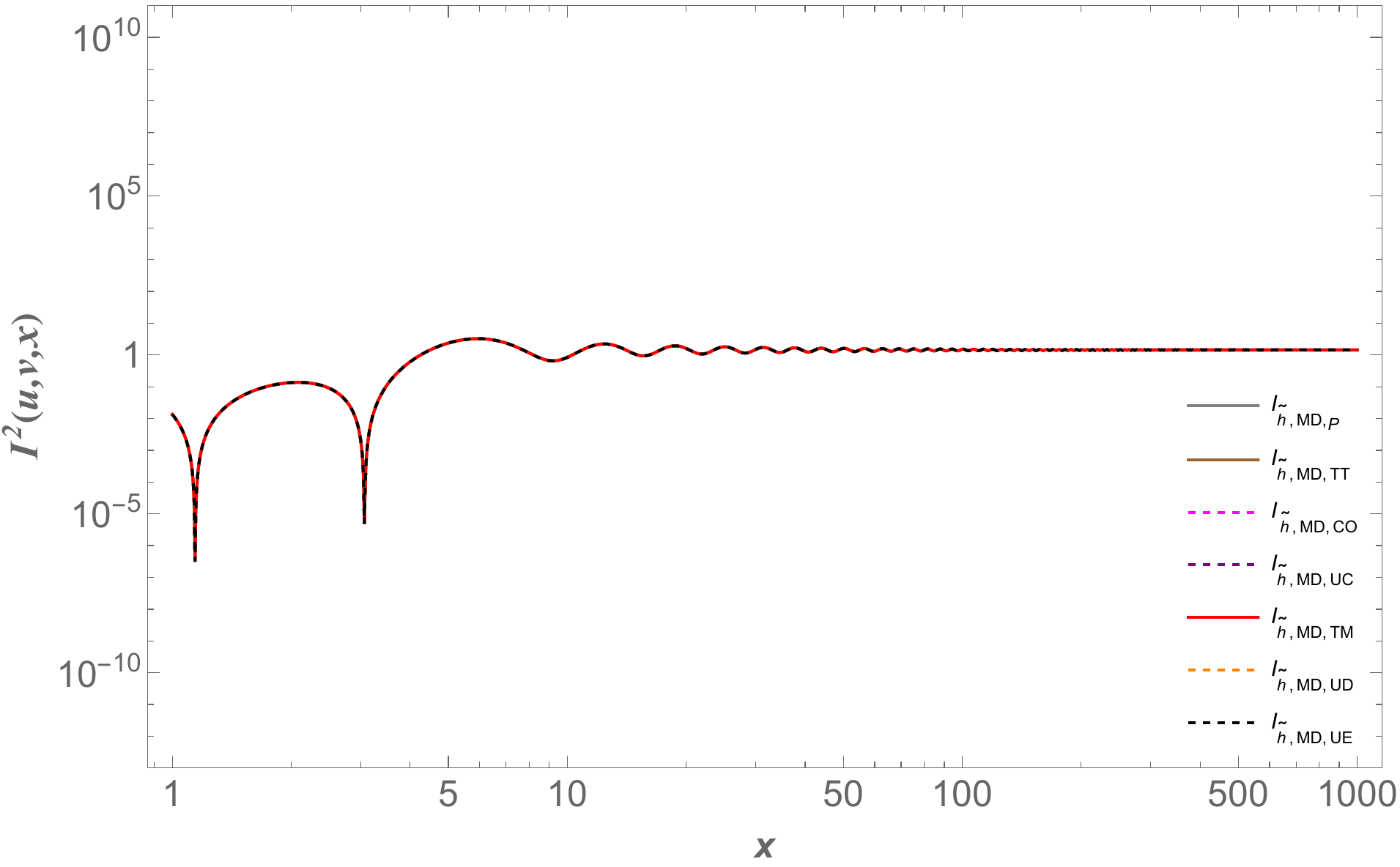}%
    \qquad
    \includegraphics[width=.46\linewidth]{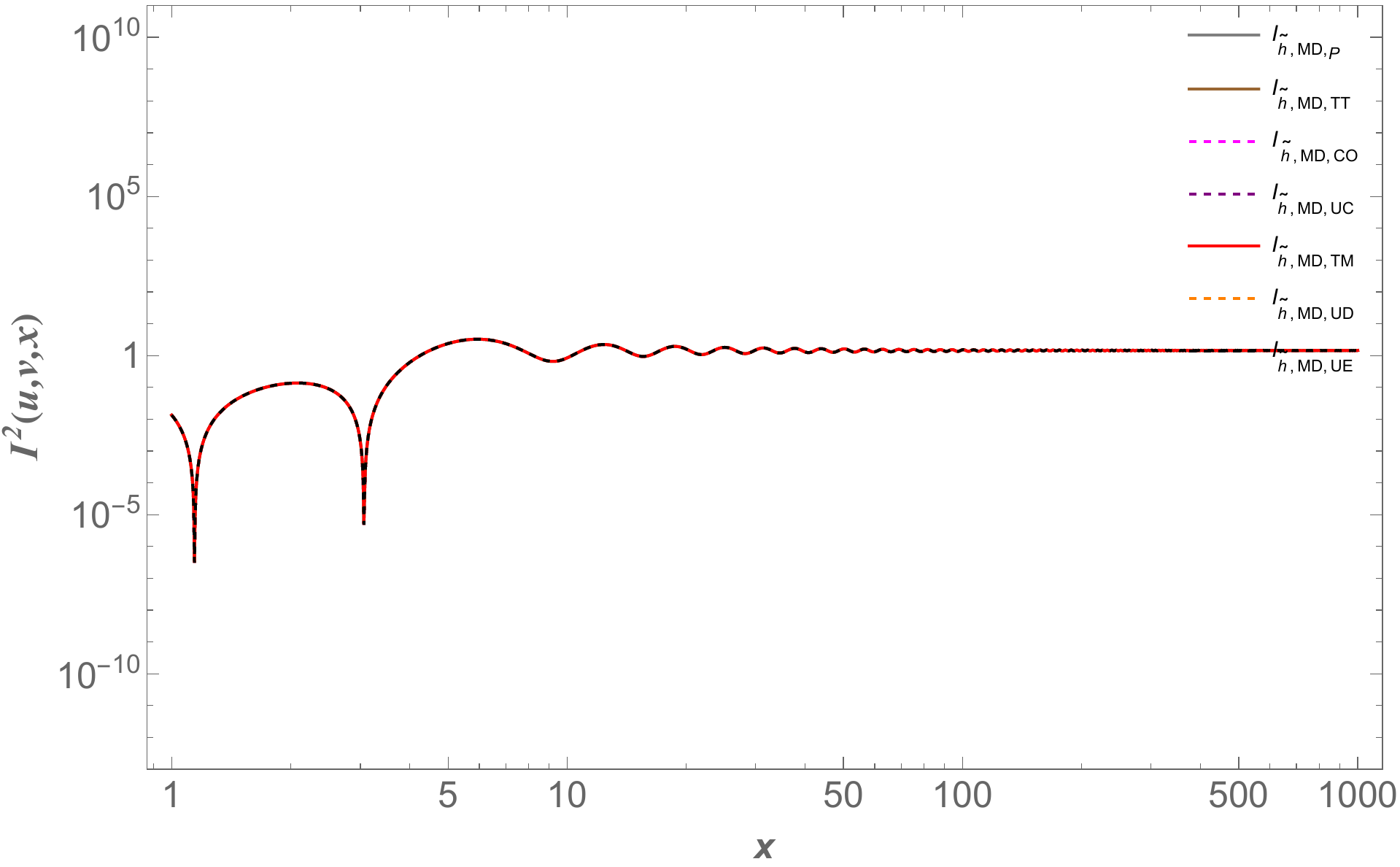} %
    \caption{(Color online) The evolution of $\lvert I_{\tilde{h},\,\text{MD},P}\rvert$, $\lvert I_{\tilde{h},\,\text{MD},\text{TT}}\rvert$, $\lvert I_{\tilde{h},\,\text{MD},\text{CO}}\rvert$,
    $\lvert I_{\tilde{h},\,\text{MD},\text{UC}}\rvert$, $\lvert I_{\tilde{h},\,\text{MD},\text{TM}}\rvert$, $\lvert I_{\tilde{h},\,\text{MD},\text{UD}}\rvert $, and $\lvert I_{\tilde{h},\,\text{MD,UE}}\rvert$ is  shown to be constant at the subhorizon limit in all gauges. Here, we show kernels  with a constant factor 6/5. {{Left:}} we take $u=v=1.$ {{Right:}} we let $u=1$ \text{and}  $v=0.1$.}%
    \label{fig:convergent.MD}%
\end{figure*}

This section compares the kernel functions in the Poisson gauge, the TT gauge, the comoving orthogonal gauge, the uniform curvature gauge, the total matter gauge, the uniform density gauge, and the uniform expansion gauge during RD and MD. We show that they all lead to the same gauge-independent kernel functions. Thus, the energy density, $\Omega_{\text{GW}}$, of SIGWs should be the same in seven gauge choices. To be precise, in the following subsections,  we show the evolution of our results of the kernel functions in Figures \ref{fig:divergentRD} and \ref{fig:convergentRD} in RD, and  Figures \ref{fig:divergent.MD}-\ref{fig:convergent.MDafterbarringfactor}  in MD, respectively.
\subsection{Comparison among the kernel functions during RD}
	First, during RD,  we compare the behavior of $I(u, v, x)$ obtained in each gauge and their differences for the finite value of $x$. Examples of the kernel functions' time evolution for the given sets of $u=v=1$ and $u=1$, $v=0.1$ in seven gauges are shown in Figures \ref{fig:divergentRD} and \ref{fig:convergentRD}. In these figures, $x=k\eta$ can be analyzed as the time parameter in the unit in which the horizon entry of the tensor perturbation occurs at $x=1$.

Figure \ref{fig:divergentRD} shows the comparison of the evolution of $I(u,v,x)$ in
 the Poisson and six other gauges.
 As seen in this figure, before the source perturbations enter the horizon $x\ll1$, the induced perturbations remain almost constant but later start oscillating in growing or decaying modes. At the late time $x\gg1$, all of the secondary perturbations in Figure \ref{fig:divergentRD} oscillate,
and the amplitudes of their oscillations are in growing or decaying modes. This figure compares the kernel functions in the Poisson gauge with those of the six other gauges. Here, in Figure \ref{fig:divergentRD}, we show that $I_{\text{RD}}(u,v,x)$ of the tensor perturbations in six other gauges are divergent as $x\to \infty$, while the one in the Poisson gauge tends to converge. The relationship of this result was presented in refs. \cite{Lu:2020diy, Hwang:2017oxa, Tomikawa:2019tvi}, which has a flaw during RD. Also, it indicates that the tensor perturbations are gauge-dependent. Besides, the literature claims that  the physical observable energy density should not be gauge dependent.

Therefore,  we have tried to fix this discrepancy by introducing a counter term in eq. \eqref{coneqn}  and found a gauge-independent kernel function in contrast to refs. \cite{Hwang:2017oxa, Lu:2020diy, Tomikawa:2019tvi}.
In addition, the counter terms in the TT gauge, the comoving orthogonal gauge, the uniform curvature gauge, the total matter gauge, the uniform density gauge, and the uniform expansion gauge are not trivial.
Figure \ref{fig:convergentRD} shows the evolution of the gauge independent kernels $I_{\Tilde{h},\,\text{RD}}(u,v,x)$ in the Poisson and the six other gauges. After removing the discrepancy issues, we find that the physical observable $\Omega_\text{GW}$ is gauge-independent.
\subsection{Comparison among the kernels during MD}
\label{sec.3MDcase}
 Finally, we discuss comparing the kernel functions in MD.
First, we show the evolution of the kernels with the constant term $6/5$ in Figures \ref{fig:divergent.MD} and \ref{fig:convergent.MD}.
In Figure \ref{fig:divergent.MD},  we show the behavior of the evolution of $I_{\text{MD}}({u,v,x})$  at $u=v=1$ and $u=1$ and $v=0.1$, respectively.

In the left panel, we can see that the kernel functions  $I_{\text{}\text{MD},\text{CO}},I_{\text{MD},\text{UC}},I_{\text{MD},\text{TM}},I_{\text{RD},\text{UD}}$ start to grow, and $I_{\text{MD},P},I_{\text{MD},\text{TT}}$, and $I_{\text{MD,UE}}$ are going to be constant as the secondary induced perturbations enter the horizon ($x\simeq1$). While in the right panel, one can see that only $I_{\text{MD},P}$, and $I_{\text{MD,UE}}$  are going to be constant, and the others start to grow as the secondary induced perturbations enter the horizon.

From these two panels, it can be seen  that the values at $x\gg1$ do not depend on $u$ and $v$ extensively except for $I_{\text{}\text{MD},\text{CO}}$, and $I_{\text{MD},\text{TT}}$. As for $I_{\text{}\text{MD},\text{CO}}$, and $I_{\text{MD},\text{TT}}$, these have different behavior for $v=1$ and $v=0.1$ before the source perturbations enter the horizon ( $x\ll1$) and at the late time  ($x\gg1$).
We can see that the behavior of these kernels is not the same as those in an RD era. From these observations, one can deduce that the behaviors of the
secondary perturbations induced by first perturbations at $(x\gg1)$ are distinct except in Poisson and uniform expansion gauges.

\begin{figure*}[!t]
\centering
 \includegraphics[width=.46\linewidth]{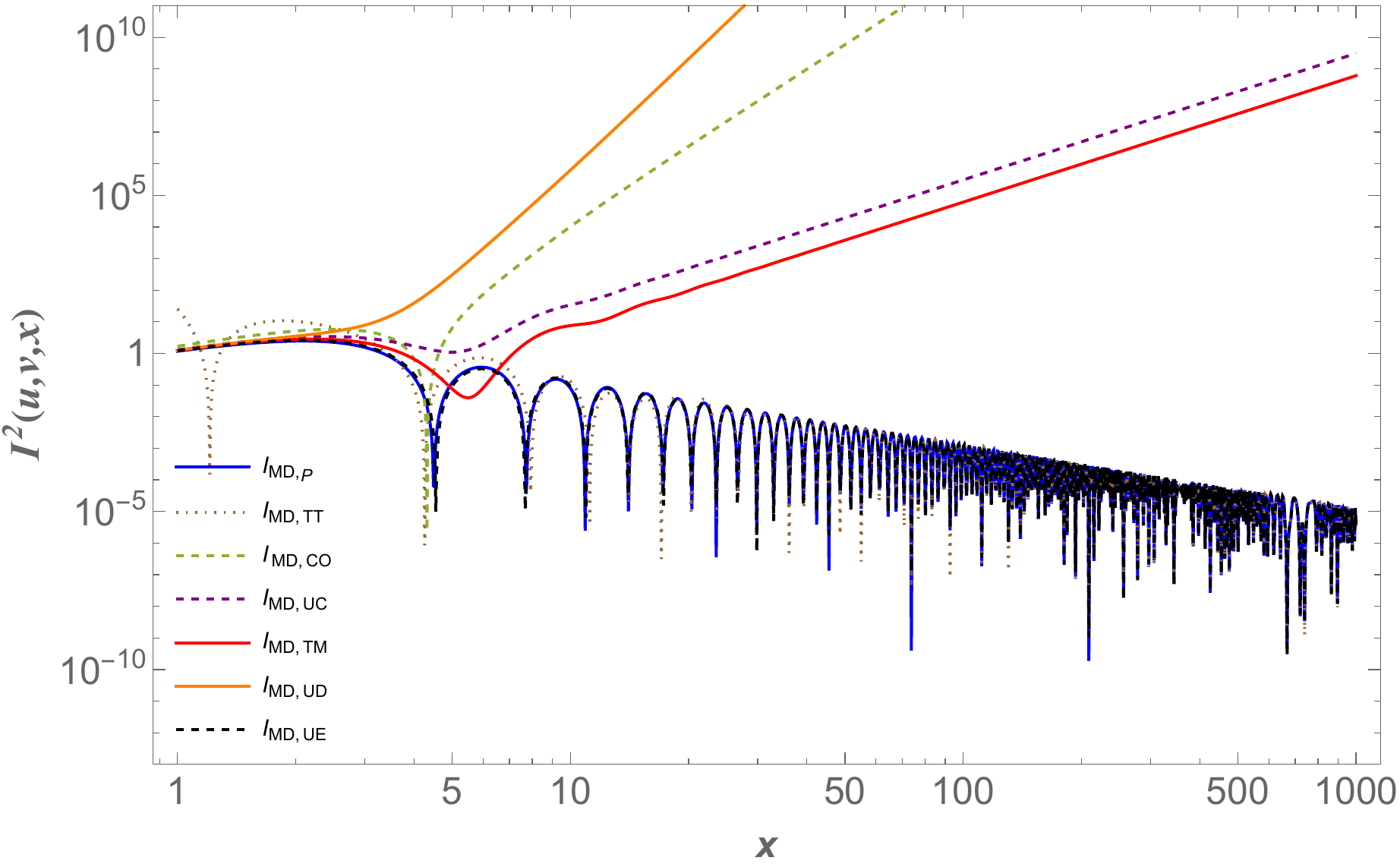} \qquad
    \includegraphics[width=.46\linewidth]{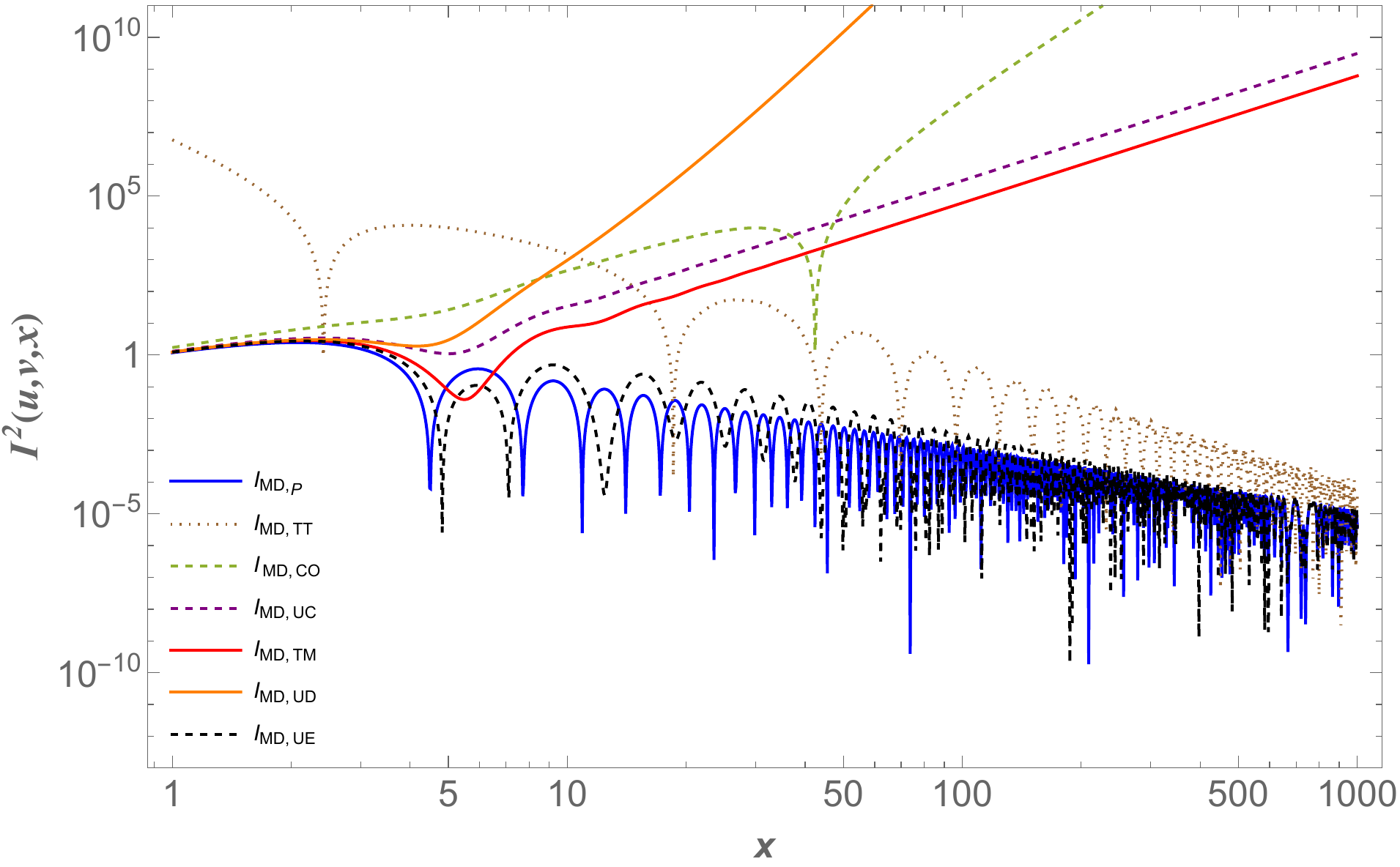}%
    \caption{(Color online) The evolution of $\lvert I_{\text{MD},P}\rvert$, $\lvert I_{\text{MD},\text{TT}}\rvert $, $\lvert I_{\text{}\text{MD},\text{CO}}\rvert$, $\lvert I_{\text{MD},\text{UC}}\rvert$, $\lvert I_{\text{MD},\text{TM}}\rvert $, $\lvert I_{\text{MD},\text{UD}}\rvert$,  and $\lvert I_{\text{MD,UE}}\rvert$ is  shown to be divergent in different gauges. Here, we show kernels after barring the factor 6/5. {{Left:}} we take $u=v=1.$ {{Right:}} we let $u=1$ {and}  $v=0.1$.}    \label{fig:divergent.MDbarringfactor}
\end{figure*}

\begin{figure*}[!t]
\centering\vspace*{2mm}
  \includegraphics[width=.46\linewidth]{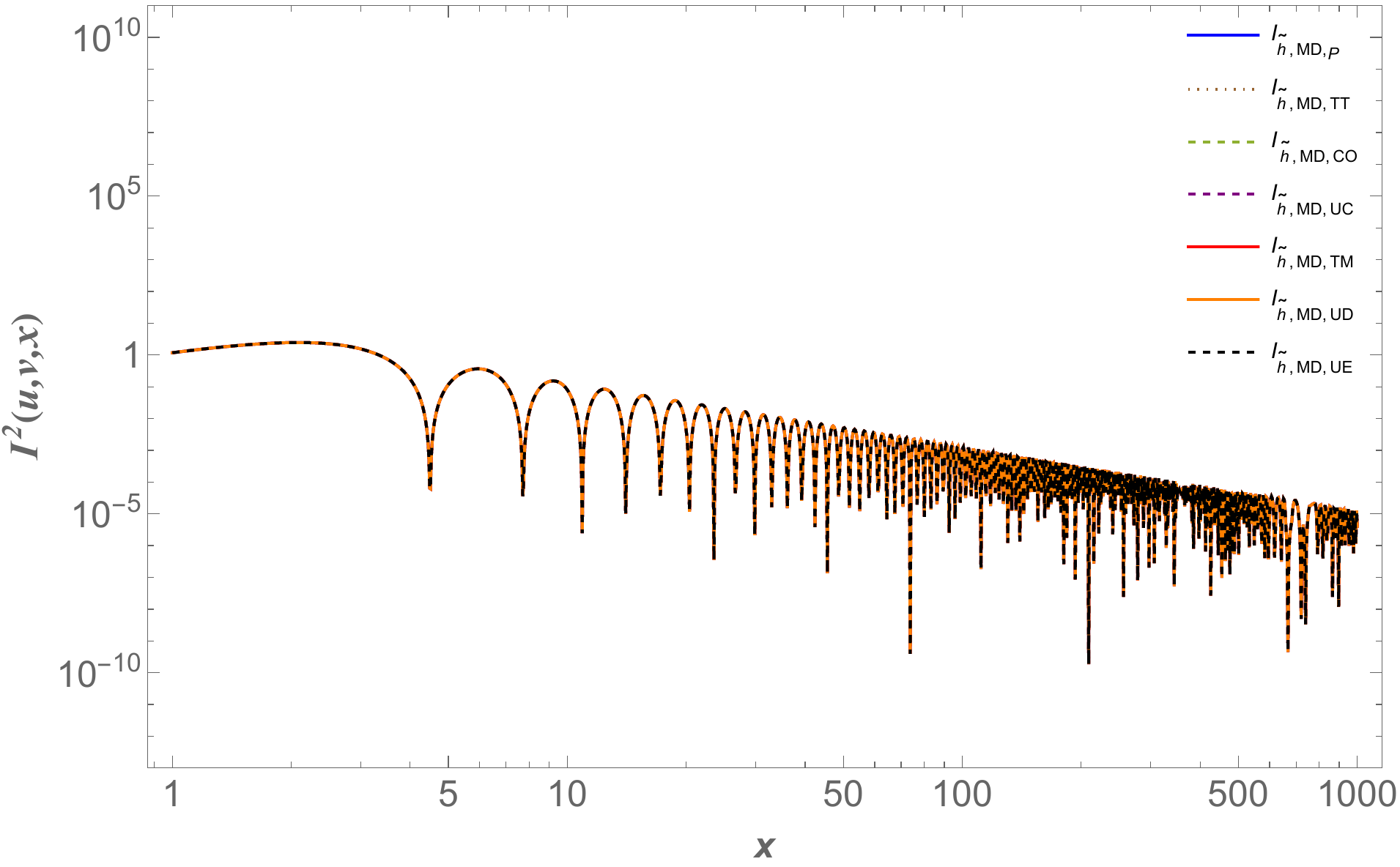} \qquad
    \includegraphics[width=.46\linewidth]{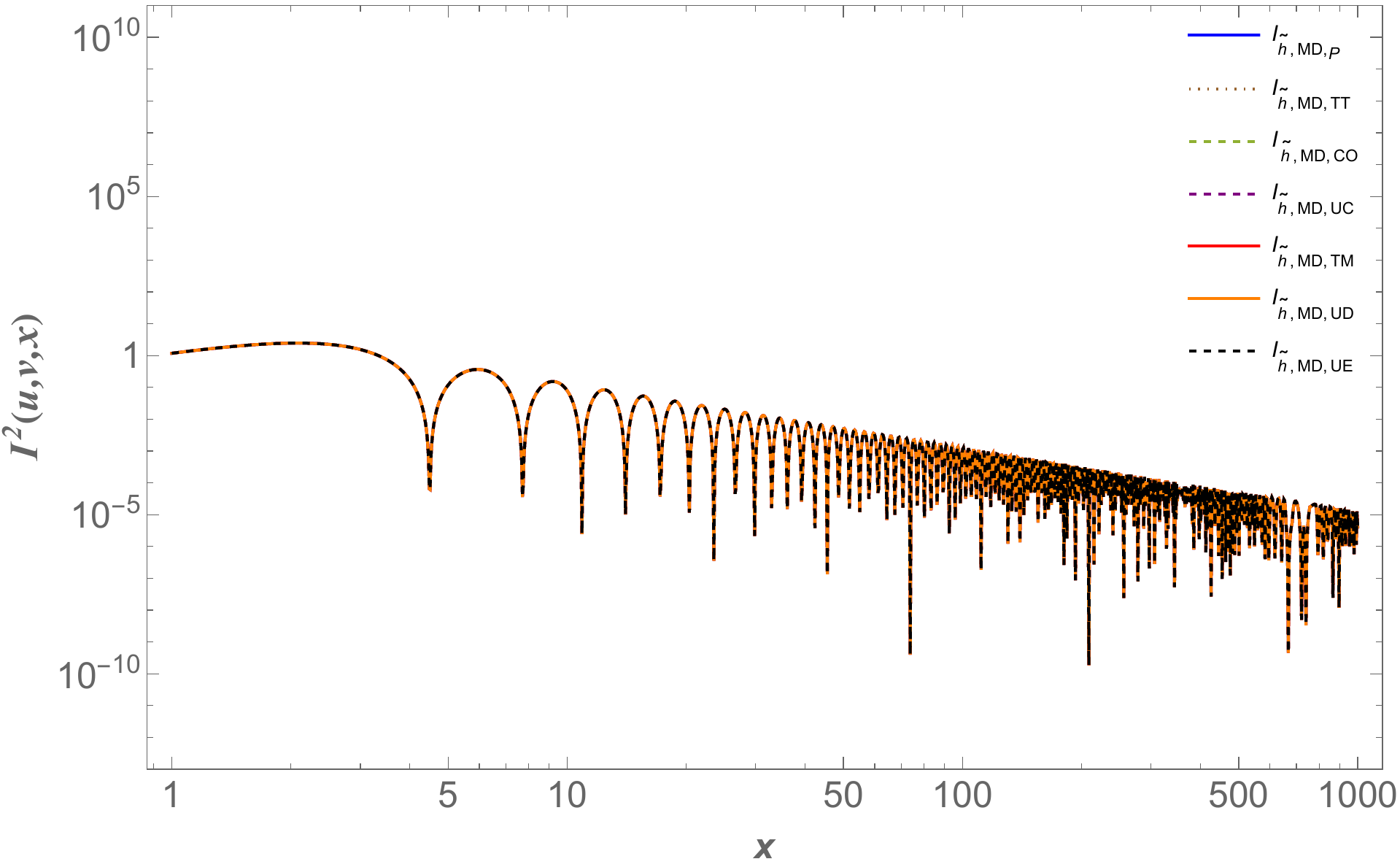}%
    \caption{(Color online) The evolution of $\lvert I_{\tilde{h},\,\text{MD},P}\rvert$, $\lvert I_{\tilde{h},\,\text{MD},\text{TT}}\rvert$, $\lvert I_{\tilde{h},\,\text{MD},\text{CO}}\rvert$,
    $\lvert I_{\tilde{h},\,\text{MD},\text{UC}}\rvert $, $\lvert I_{\tilde{h},\,\text{MD},\text{TM}}\rvert$,
    $\lvert I_{\tilde{h},\,\text{MD},\text{UD}}\rvert $,  and $\lvert I_{\tilde{h},\,\text{MD,UE}}\rvert$ is shown to be convergent in all of the gauges. Here, we show kernels after barring the factor 6/5. {{Left:}} we take $u=v=1.$ {{Right:}} we let $u=1$ and $v=0.1$.}
        \label{fig:convergent.MDafterbarringfactor}%
\end{figure*}

Moreover, in Figure \ref{fig:convergent.MD}, after considering the terms (except the single term like $x^n$ or $1/x^n$) in six other gauges, like in the Poisson gauge, we get the same behavior in the evolution of the kernel functions in all seven gauges, which is actually a constant behavior in $x\gg1$.
 We deduce that the behavior of secondary perturbations at $(x\gg1)$ is the same, and all the kernels are almost constant as $x\to \infty$. As discussed in the previous subsection, these behaviors are not identical to those in an RD universe. Because we included the constant terms in each gauge, they do not represent the GWs oscillations as shown in the previous subsection.

Finally, we compare the kernels after barring the factor 6/5. Specifically, we show the evolution of our results for the kernel functions in Figures \ref{fig:divergent.MDbarringfactor} and \ref{fig:convergent.MDafterbarringfactor}.
In Figure \ref{fig:divergent.MDbarringfactor}, in the left panel, it is to be noted that kernel functions of tensor perturbations in the comoving gauge, the comoving orthogonal gauge, the uniform curvature gauge, and the uniform density gauge start to grow on superhorizon scales $(x\simeq1)$ and tend to be divergent as $x\to \infty$, while those in the Poisson, the TT, and the uniform expansion gauges tend to converge at the lat time limit $(x\gg1)$.
In the MD era, we also used
eq. \eqref{coneqn}  and found a gauge-independent kernel functions $I_{\Tilde{h},\, \text{MD}}(u,v,x)$ in contrast to refs. \cite{Hwang:2017oxa, Tomikawa:2019tvi, Ali:2020sfw}.
\section{Discussion and concluding remarks}\label{conclus}
In this paper, we have reconsidered the SIGWs in the late-time limit in different popular gauges. In particular, we have dealt with the second-order tensor perturbations generated by the linear scalar perturbations in an expanding spacetime containing either RD or MD.
 We have tried to address the discrepancies of the previous studies by introducing
a counter term \eqref{counterterm} to remove the fictitious terms in the secondary tensor perturbations.  We have shown that the late-time (e.g., observable) GWs investigated in seven different gauges coincide with each other, in contrast to refs. \cite{Hwang:2017oxa, DeLuca:2019ufz, Lu:2020diy, Tomikawa:2019tvi, Ali:2020sfw}.
In this work, we have explicitly evaluated the gauge-independent kernel functions, which uniquely examined the energy density of SIGWs in different gauges. Moreover, the evolution of the transfer functions is also presented.

	On the other hand, according to refs. \cite{Hwang:2017oxa, DeLuca:2019ufz, Lu:2020diy, Tomikawa:2019tvi, Ali:2020sfw}, the secondary tensor perturbations could be different in seven different gauges, even in the subhorizon limit. One can find that the difference between the Poisson gauge and the six other gauges comes from the extra terms like  $\cos \left(\frac{u\pm v x}{\sqrt{3}}\right)$ or $\sin \left(\frac{u\pm v x}{\sqrt{3}}\right)$, and $x^n$ or $1/x^n$; see, the specific expressions of different kernels in refs. \cite{Hwang:2017oxa, DeLuca:2019ufz, Lu:2020diy, Tomikawa:2019tvi, Ali:2020sfw}. This indicates that these different results of kernels in different gauges show gauge dependence that occurs in the secondary tensor perturbations coupling with scalar perturbation, not in the GWs.
 More precisely, in the gauge independent  framework in the RD phase,
 the discrepancy appearing in different gauges in refs. \cite{Hwang:2017oxa, DeLuca:2019ufz, Lu:2020diy, Tomikawa:2019tvi}  is eliminated.
 Consequently, we have found the gauge-independent kernel functions, which uniquely determine the same energy density $\Omega_\mathrm{GW}$ of SIGWs in seven gauges.

It is to be noted that the situation is different for the secondary tensor perturbations induced by scalar perturbations in the late MD phase. In this case, the scalar perturbations continue to accompany the secondary tensor perturbations on the subhorizon limit, even in the Poisson gauge, because of the growing matter perturbations.  This kind of secondary tensor perturbation can be larger than the induced GWs during the RD phase on a large scale \cite{Baumann:2007zm}. That is why the secondary tensor perturbations easily depend on the gauge during the MD phase. For example, in ref.~\cite{Ali:2020sfw}, it is shown there in different gauges that the kernel function   $I(u,v,x)$ is different in other gauges from that in the Poisson gauge even at late time. However, it should be noted that this kind of secondary tensor perturbation is not a gravitational wave. In this work, we remove the occurring discrepancies by introducing the counter term. Here, we also find the gauge independent energy density $ \Omega_\mathrm{GW}$ of SIGWs.

Observationally, the secondary tensor perturbations are usually assumed to be GWs. However, the observational sensitivity for these GWs will be distinct from that of conventional GWs.
The secondary induced tensor perturbations may explain the signal observed by the North American Nanohertz Observatory for Gravitational Waves (NANOGrav) \cite{NANOGrav:2020gpb,Rezazadeh:2021clf}.

Finally, it is important to recognize that our analysis can be extended in various ways. In particular, we explore SIGWs further to examine the physics of the early Universe in the inflationary scenario and the associated PBHs.
Further, we need to study the gauge dependency of SIGWs to find a gauge-invariant way.
Moreover, we will explore the waveform of the energy density $\Omega_\mathrm{GW}$ and examine its relationship with the scalar power spectrum
$\mathcal{P}_\zeta^2(k)$ during RD and MD.

\Acknowledgements{Arshad Ali and Mudassar Sabir are thankful to Professor Yungui Gong for many inspiring discussions and collaborations on related topics. \\
This work was supported by the National Natural Science Foundation of China (Grant Nos. 12175105, 12147175, 12247170, 11575083, and 11565017),  the Top-notch Academic Programs Project of Jiangsu Higher Education Institutions (TAPP). }

\end{multicols}

\end{document}